\documentclass[pre,aps,
twocolumn,
superscriptaddress,floats,showpacs,10pt]{revtex4-1}
\usepackage{graphicx}
\usepackage{dcolumn}
\usepackage{bm}
\usepackage{amssymb}
\usepackage{amsmath}
\usepackage{array}
\usepackage{float}
\usepackage{geometry}
\usepackage{color}
\usepackage{amsthm}

\begin{document}

\title{Nonlinear Plasma Wave in Magnetized Plasmas}
\author{Sergei V. Bulanov}
\altaffiliation[Also at ]{Prokhorov Institute of General Physics, Russian Academy of Sciences, Moscow
119991, Russia}
\affiliation{Kansai Photon Science Institute, JAEA, Kizugawa, Kyoto 619-0215, Japan}
\author{Timur Zh. Esirkepov}
\affiliation{Kansai Photon Science Institute, JAEA, Kizugawa, Kyoto 619-0215, Japan}
\author{Masaki Kando}
\affiliation{Kansai Photon Science Institute, JAEA, Kizugawa, Kyoto 619-0215, Japan}
\author{James K. Koga}
\affiliation{Kansai Photon Science Institute, JAEA, Kizugawa, Kyoto 619-0215, Japan}
\author{Tomonao Hosokai}
\affiliation{Photon Pioneers Center, Osaka University, 2-8 Yamadaoka, Suita, Osaka 565-0871, Japan}
\author{Alexei G. Zhidkov}
\affiliation{Photon Pioneers Center, Osaka University, 2-8 Yamadaoka, Suita, Osaka 565-0871, Japan}
\author{Ryosuke Kodama}
\affiliation{Photon Pioneers Center, Osaka University, 2-8 Yamadaoka, Suita, Osaka 565-0871, Japan}
\affiliation{Graduate School of Engineering, Osaka University, 2-1 Yamadaoka, Suita, Osaka 565-0871, Japan}

\maketitle

{\bf Abstract} 
Nonlinear axisymmetric cylindrical plasma oscillations in magnetized collisionless plasmas 
are a model for the electron fluid collapse on the axis behind an ultrashort relativisically intense laser 
pulse exciting a plasma wake wave. We present an analytical description of the strongly nonlinear oscillations
showing that the magnetic field prevents closing of the cavity formed behind the laser pulse. 
This effect is demonstrated with 3D PIC simulations
of the laser-plasma interaction. An analysis of the betatron oscillations of fast electrons in the presence of the magnetic field
 reveals a characteristic ``Four-Ray Star" pattern which has been observed in the image of the electron bunch in experiments 
 [T. Hosokai, {\it et al.}, Phys. Rev. Lett. {\bf 97}, 075004 (2006)].

\section{Introduction}
Theoretical and experimental studies of electromagnetic wave propagation in magnetized plasmas are of key importance for 
a vast range of problems in space and laboratory physics \cite{VLG, FFC}. The interest towards relativistically strong plasma waves stems both from 
astrophysical applications and from high field science, i. e. from relativistic plasmas research \cite{MTB}. 
Relativistic plasma waves can be generated as 
wake waves in a plasma behind an ultrashort laser pulse \cite{TD, ESW} or
by bunches of ultra-relativistic electrons \cite{PC},  providing extremely strong electric fields used for the acceleration of
high energy charged particles. Electron acceleration by the plasma wake field above energies of GeV-ies 
has been demonstrated in a number of experiments \cite{LWFA}.
Physical processes associated with the nonlinear wake wave play a key role in high-harmonic generation
\cite{HHG} and in many other aspects of laser-plasma physics \cite{MTB}, including the frequency 
up-shifting and intensification of the electromagnetic radiation \cite{FM, BUFN}.

An important property of a nonlinear wake wave is that when its amplitude exceeds the wavebreaking threshold 
the wake breaks transferring its energy to the plasma electrons, thus, 
on one hand, providing a robust mechanism of electron injection to the acceleration phase of the wake field \cite{INJ}
and forming regular nonlinear structures moving with relativistic velocity \cite{BUFN}  and, on the other hand,
 imposing a constraint on the achievable wake wave amplitude \cite{AP, TWB}. 
 A tightly focused and sufficiently intense laser pulse excites a wake 
 wave in the regime, where in addition to a longitudinal push the laser pulse expels electrons 
 also in transverse direction \cite{TWB, PMtV, BW1, BW2}, forming a cavity void of
electrons in the first period of the wake wave. The parameters of the cavity in the electron density behind the
laser pulse can be deduced from the fact that all the electrons for a 
sufficiently intense laser pulse are pushed aside resulting in a multiflow motion
owing to the collisionless nature of the low-density
plasma \cite{BW1,BW2}. The region void of electrons is positively charged
and attracts unperturbed electrons on the side. 
In the near axis region, where the cavity is closing, the electron fluid dynamics 
can be modeled by that of an axisymmetric cylindrical Langmuir wave. Nonlinear cylindrical and spherical Langmuir waves 
have been considered in Refs. \cite{DGBSS}, where it was shown that the finite amplitude electrostatic waves break
due to nonlinear effects and the non-planar geometry.

Below we report the results of analytical investigations and computer simulations of nonlinear Langmuir oscillations 
in a plasma with a relatively strong homogeneous axial magnetic field. There are several mechanisms of strong magnetic field generation in the laser matter interaction.
Large amplitude magnetic fields are generated by temperature gradients near the laser focal spot, by the hot electron currents driven
by an ultraintense laser pulse, in the plasma jets induced by them, and by the compression of thin shells by the laser pulses \cite{LMF}. 
Axial magnetic fields can be generated by circularly polarized laser radiation via the Inverse Faraday Effect \cite{IFE}.
Strong magnetic fields can change the 
 whole scenario of the laser plasma interaction \cite{GAA} and, in particular, 
 they can modify the charged particle acceleration by electrostatic waves \cite{STRN, OTHMF} and the injection \cite{BINJ}. 
 The magnetic field with required for the injection enhancement symmetry can be produced inside capillary plasma targets 
 used for laser pulse guiding \cite{CAPILL}.
 Imposed external homogeneous magnetic fields can significantly improve the quality and stability of the laser wakefield accelerated electrons \cite{THBf}.
 In the case of electrostatic waves their frequency lies in the vicinity of the upper hybrid resonance \cite{FFC}. 
 Upper hybrid wake wave generation by laser light 
 has been investigated in Refs. \cite{UHWW}. Their properties have also attracted attention with regard to the acceleration 
 of ultra high energy cosmic rays in the astrophysical environment \cite{UHECR}.
 
 When an extraordinary electromagnetic wave (whose electric field is perpendicular to the external magnetic field, ${\bf E}\perp {\bf B}_0$) 
 of small amplitude propagates in the direction transverse to the magnetic field, ${\bf k} \perp {\bf B}_0$, 
 its frequency, $\omega$, and wave number, $k$,  are related by the dispersion equation
 \begin{equation}
 (\omega^2-\omega^2_{pe})(\omega^2-k^2 c^2-\omega^2_{pe})=\omega^2_{Be}(\omega^2-k^2 c^2).
 \label{eq-DEq}
 \end{equation}
Here $\omega_{pe}=\sqrt{4\pi n e^2/m_e}$ and $\omega_{Be}=e B_0/m_e c$ are the Langmuir frequency and the electron Larmor frequency, respectively, 
with $n$, $B_0$, $e$, $m_e$ and $c$ being the plasma density, the external magnetic field strength, electron charge and mass, and speed of light in vacuum. 
The ions are assumed to be at rest.
When the magnetic field vanishes, the dispersion equation (\ref{eq-DEq}) describes two independent waves: 
the electrostatic longitudinal (${\bf k} || {\bf E}$) Langmuir wave with the frequency equal to $\omega=\omega_{pe}$, 
and the electromagnetic transverse (${\bf k} \perp {\bf E}$) wave, for which the frequency and wave number are related as $\omega=\sqrt{k^2 c^2+\omega^2_{pe}}$. 
In a plasma with a finite external magnetic field, $B_0\ne 0$, 
the extraordinary electromagnetic wave has both components of the electric field parallel and 
perpendicular to the magnetic field.

\begin{figure}[tbph]
\centering
\includegraphics[width=6cm,height=2.5cm]{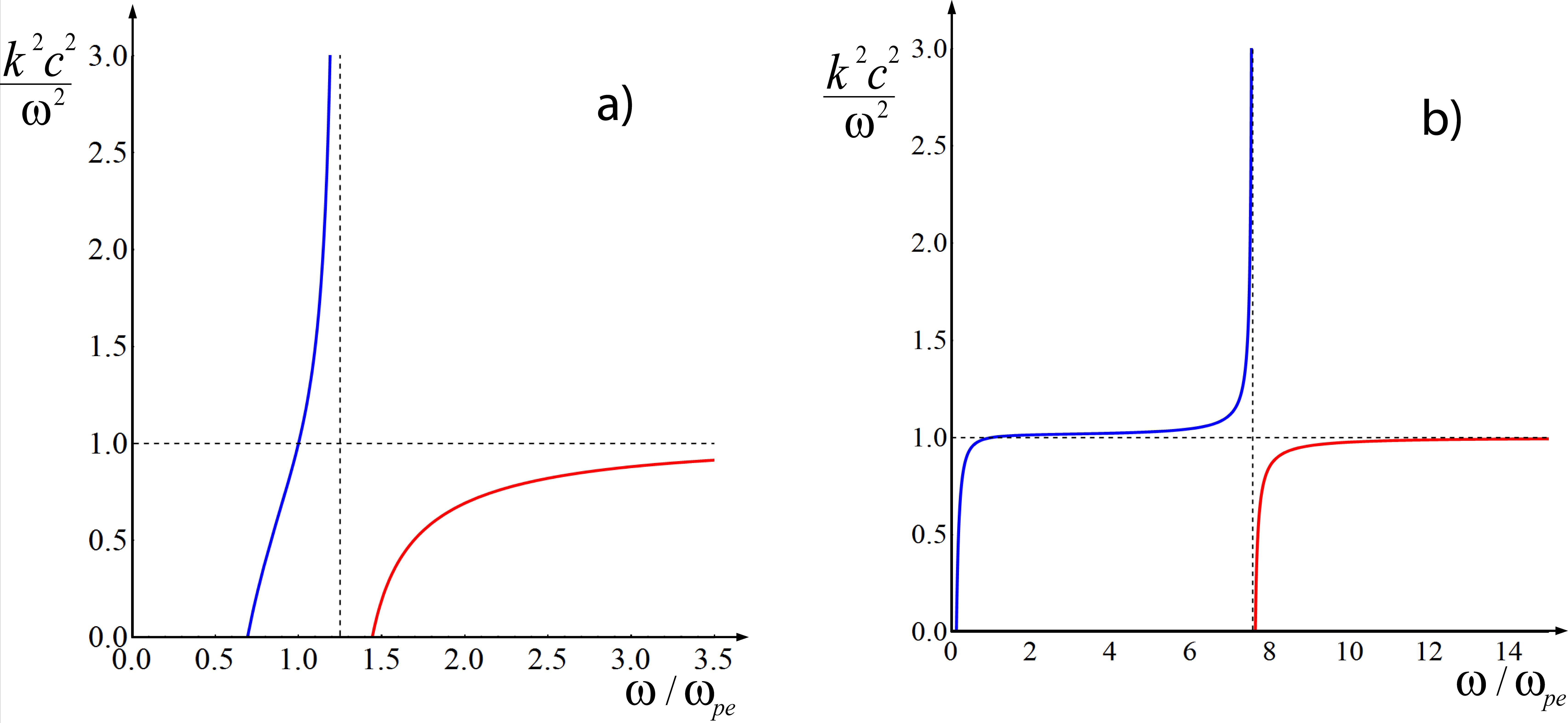}
\caption{Square of the refraction index, $N^2=k^2 c^2/\omega^2$, as a function of the wave frequency normalized by  
$\omega_{pe}$: a) $\omega_{Be}/\omega_{pe}$=7.5, 
b) $\omega_{Be}/\omega_{pe}$=0.75.}
\label{fig1}
\end{figure}

In order to illustrate the extraordinary electromagnetic wave dispersion properties, 
in Fig. \ref{fig1} we plot the square of the refraction index, $N^2=k^2 c^2/\omega^2$, equal to
 \begin{equation}
N^2=\frac{\omega^4-\omega^2(2\omega^2_{pe}+\omega^2_{Be})+\omega^4_{pe}}{\omega^2\left(\omega^2-\omega^2_{pe}-\omega^2_{Be}\right)}
 \label{eq-K}
 \end{equation}
as a function of the wave frequency for relatively large and 
small magnetic field. As we see the refraction index tends to infinity at the upper hybrid resonance, where the wave frequency is equal to 
 \begin{equation}
\omega_{UH}=\sqrt{\omega^2_{pe}+\omega^2_{Be}}.
 \label{eq-omUH}
 \end{equation}
 
In the vicinity of the upper hybrid resonance, which corresponds to the limit $k\gg \omega_{UH}/c$, the wave is electrostatic, i.e. one 
can neglect the displacement current effects. 

\section{Planar Upper Hybrid Wave}

\subsection{Relativistic upper hybrid oscillations}
In order to compare the properties of a nonlinear upper hybrid wave in the planar and axial symmetric geometries we at first consider a planar electrostatic wave 
propagating perpendicular to a constant homogeneous magnetic field directed along the z-axis, ${\bf B}=B_0 {\bf e}_z$, 
when the functions under consideration depend on time and the coordinate $x$.
The electron hydrodynamics equations and Maxwell's equations yield for the electron density $n$, electron momentum ${\bf p}=p_x {\bf e}_x+p_y {\bf e}_y$ and 
the electric field $E_x$:
 \begin{equation}
\partial_t n+\partial_x (n v_x)=0,
 \label{eq-Cont1D}
 \end{equation}
  \begin{equation}
\partial_t p_x+v_x\partial_x p_x=-e E_x-\frac{e}{c} v_{y} B_0,
 \label{eq-px1D}
 \end{equation}
 \begin{equation}
\partial_t p_{y}+v_x\partial_x p_{y}=\frac{e}{c} v_x B_0,
 \label{eq-py1D}
 \end{equation}
  \begin{equation}
\partial_t E_x+v_x\partial_x E_x=4\pi n_0 e v_x,
 \label{eq-Ex1D}
 \end{equation}
 where $v_x=c p_x/\sqrt{m_e^2c^2+p_x^2+p_y^2}$ and $v_y=c p_y/\sqrt{m_e^2c^2+p_x^2+p_y^2}$. 
 The nonrelativistic limit has been analyzed in Ref. \cite{RD}.
 
Changing from Euler coordinates $x,t$ to Lagrange variables $x_0,t$ we obtain the relationship between $x$ and $x_0$: 
 \begin{equation}
x=x_0+\xi(x_0,t)
 \label{eq-EuLa}
 \end{equation}
 with $\xi(x_0,t)$ being the displacement of the electron fluid element from its initial position, $x_0$. 
 The electron velocity is equal to $v_x=\dot \xi$. Here and below a dot denotes a differentiation with the respect 
 to time in the Lagrange variables.
 
 In the Lagrange coordinates the system of equations (\ref{eq-px1D} - \ref{eq-Ex1D}) takes the form
  \begin{equation}
\dot p_x=-e E_x-\frac{e}{c} v_{y} B_0,
 \label{eq-px1DLa}
 \end{equation}
 \begin{equation}
\dot p_{y}=\frac{e}{c} v_x B_0,
 \label{eq-py1DLa}
 \end{equation}
  \begin{equation}
\dot E_x=4\pi n_0 e v_x.
 \label{eq-Ex1DLa}
 \end{equation}
 
 The solution to the continuity equation for the electron density reads: $n(x_0,t)=n_0/|\partial_{x_0}x|$.
 Integration of Eqs. (\ref{eq-py1DLa}) and (\ref{eq-Ex1DLa}) 
 yields $p_{y}=(e B_0/c) \xi+{\mathbb P}_y (x_0)$ and $E_x=4\pi n_0 e \xi+{\mathbb E}_x(x_0)$, which shows that $p_y$ and $E_x$
 are related to each other via 
  \begin{equation}
\frac{cp_y}{eB_0}+\frac{E_x}{4\pi n_0e}=\frac{c{\mathbb P}_y (x_0)}{eB_0}+\frac{{\mathbb E}_x(x_0)}{4\pi n_0e}.
 \label{eq-ExPy1DLa}
 \end{equation}
 Here the functions ${\mathbb P}_y (x_0)$ and ${\mathbb E}_x(x_0)$ are determined by the initial conditions.
 Assuming ${\mathbb P}_y (x_0)=0$ and ${\mathbb E}_x(x_0)=0$ and substituting them in the r.h.s. of Eq. (\ref{eq-px1DLa})
 we find that this equation can be written in the Hamiltonian form with the Hamilton function 
 $$ {\cal H}(\xi,p_x)=\qquad \qquad \qquad \qquad \qquad \qquad \qquad \qquad $$
  \begin{equation}
\sqrt{m_e^2c^4+(e B_0)^2\xi^2+p_x^2c^2}+2\pi n_0 e^2 \xi^2.
 \label{eq-Ham1DLa}
 \end{equation}
 Since the Hamilton function (\ref{eq-Ham1DLa}) does not depend explicitly on time, the conservation of
 ${\cal H}(\xi,p_x)=h(x_0)$ gives a relationship between the electron momentum, $p_x$, and the displacement, $\xi$
 along the trajectory determined by $h(x_0)$. The maximal electron momentum, $p_{x,m}=m_ec\sqrt{\gamma_m^2-1}$, with $\gamma_m=h/m_e c^2$, 
 and the displacement 
 amplitude, $\xi_m$, are related to each other as 
 $$\xi_m=\qquad \qquad \qquad \qquad \qquad \qquad \qquad \qquad $$
\begin{equation}
\sqrt{2}\sqrt{\varepsilon_B^2 +\gamma_m -
\sqrt{\varepsilon_B^4+2\gamma_m\varepsilon_B^2+1}}.
\label{eq-PmXim}
 \end{equation}
 Here and below 
 \begin{equation}
 \varepsilon_B=\frac{\omega_{Be}}{\omega_{pe}}.
 \label{eq-epsB}
 \end{equation}
 The value of $h(x_0)$ determines the trajectory of Hamiltonian system ({\ref{eq-px1DLa}). Its phase portrait, 
 which coincide here with the contours of constant values of the Hamiltonian (\ref{eq-Ham1DLa}), is presented in Fig. \ref{fig2}.
 \begin{figure}[tbph]
 \centering
\includegraphics[width=6cm,height=6cm]{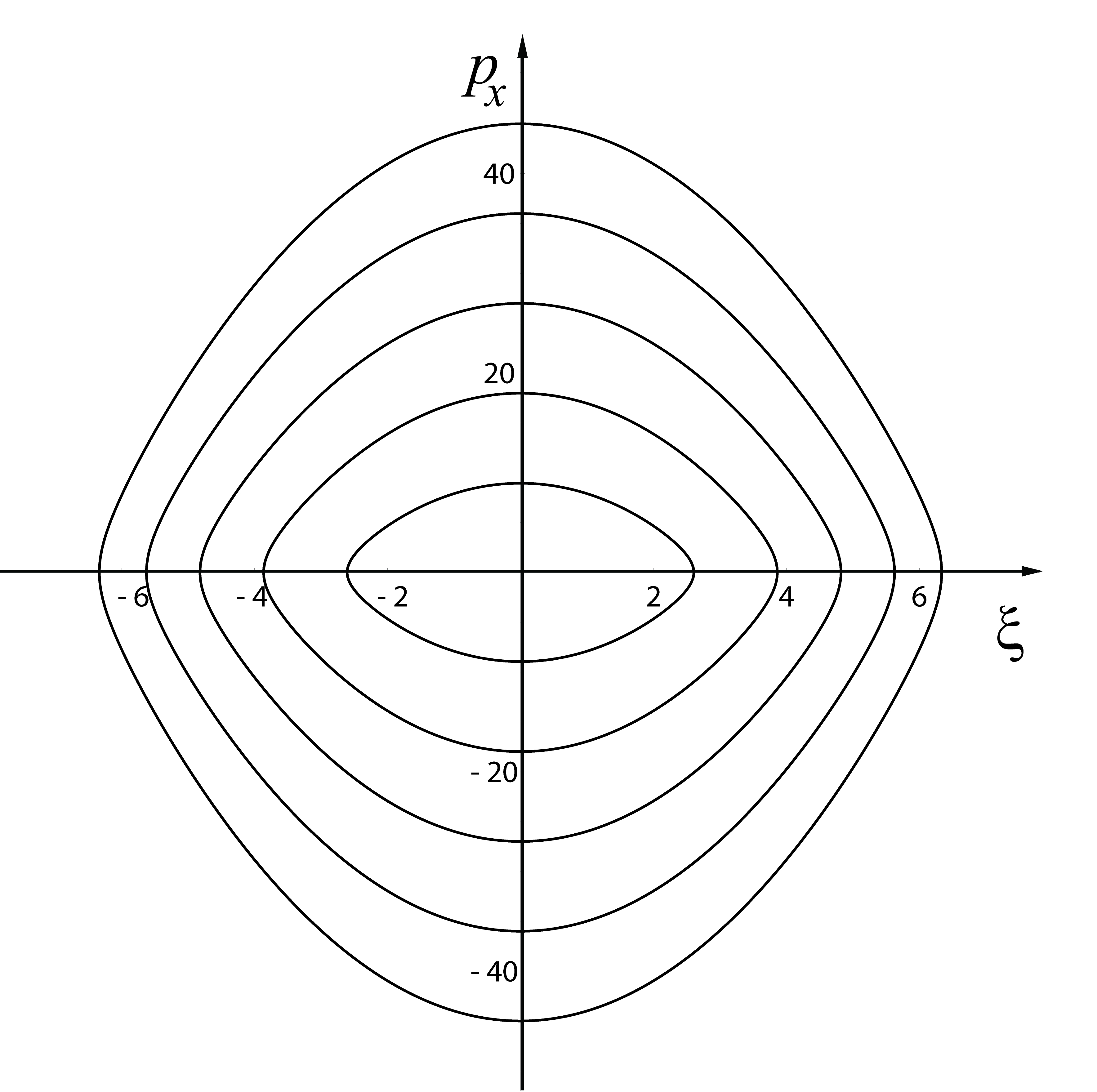}
\caption{Contours of constant values of the Hamiltonian (\ref{eq-Ham1DLa}) for $\varepsilon_B=0.75$, $\omega_{pe}=1$ and $h=9,18,27,36,45$.}
\label{fig2}
\end{figure}
 
 \subsection{Cold wave breaking limit for an upper hybrid wave travelling with constant velocity}
We assume that the wave propagates with a constant phase velocity $v_{ph}=c\beta_{ph}$ (the group velocity of the upper hybrid wave in a cold plasma vanishes)
in a cold plasma with immobile ions and homogeneous density $n_0$.
In this case all dependent variables are functions of the variable $X=x-c\beta_{ph} t$. From Eqs. (\ref{eq-Cont1D} -\ref{eq-Ex1D}) we obtain
 \begin{equation}
n=n_0 \frac{\beta_{ph} \sqrt{m_e^2c^2+p_x^2+p_y^2}}{\beta_{ph} \sqrt{m_e^2c^2+p_x^2+p_y^2}-p_x},
 \label{eq-n1DX}
 \end{equation}
   \begin{equation}
p'_x=-\frac{e}{c} \frac{E_x \sqrt{m_e^2c^2+p_x^2+p_y^2}+B_0 p_{y}}{\beta_{ph} \sqrt{m_e^2c^2+p_x^2+p_y^2}-p_x},
 \label{eq-px1DX}
 \end{equation}
 \begin{equation}
p'_y=-\frac{e}{c} \frac{B_0 p_{x}}{\beta_{ph} \sqrt{m_e^2c^2+p_x^2+p_y^2}-p_x},
 \label{eq-py1DX}
 \end{equation}
 \begin{equation}
E'_x=4\pi n_0 e \frac{p_{x}}{\beta_{ph} \sqrt{m_e^2c^2+p_x^2+p_y^2}-p_x}.
 \label{eq-Ex1DX}
 \end{equation}
 Here and below a prime denotes a differentiation with respect to the variable $X$.
 From Eqs. (\ref{eq-py1DX}) and (\ref{eq-Ex1DX}) we see that the relationship between the y-component of the electron momentum and the electric field:
  \begin{equation}
\frac{E_x}{4\pi n_0 e}-\frac{p_y c}{e B_0}={\rm constant}.
 \label{eq-EB0}
 \end{equation}
 
 If the constant on the r.h.s. of Eq. (\ref{eq-EB0}) vanishes, equations (\ref{eq-px1DX}) and (\ref{eq-Ex1DX}) can be cast in the form
 \[\left(c^2\frac{\beta_{ph} \sqrt{m_e^2c^2+p_x^2+p_y^2}-p_x}{\omega^2_{pe}\sqrt{m_e^2c^2+p_x^2+p_y^2}+\omega^2_{Be}m_e c}p'_x\right)'=\]
   \begin{equation}
 -\frac{p_x}
 {\beta_{ph} \sqrt{m_e^2c^2+p_x^2+p_y^2}-p_x}.
  \label{eq-1Deqmo}
 \end{equation}
 
  When the $x$-component of the electron velocity  becomes equal to the phase velocity of the
wake wave, i.e. $v_x=c\beta_{ph}$, the denominator on the r.h.s. of Eq. (\ref{eq-1Deqmo}) becomes equal to zero. This 
singularity corresponds to the wave breaking (for details see Refs. \cite{AP, TBI}) when the electron density and the gradients 
of all the functions tend to infinity.
The singularity is reached at the maximum of the
electron momentum, $p_{br}=m_ec \beta_{ph}/\sqrt{1-\beta^2_{ph}}$. 
Here it is taken into account that due to symmetry the $y$-component of the electron 
momentum vanishes at the breaking point.
Considering the wave structure
in the vicinity of the singularity at $X=X_{br}$ we can easily find that
in the nonrelativistic limit, when $\beta_{ph}\ll 1$ and $p_{br}\approx m_e v_{ph}$, the electron momentum depends on $\delta X=X-X_{br}$ as
\begin{eqnarray}
p_x= \qquad \qquad \qquad \qquad \qquad \qquad \qquad \qquad \nonumber \\
m_e v_{vp}-m_e v_{vp}\left(\frac{9}{2}\right)^{1/3}\left(\frac{\omega_{UH} \delta X}{v_{ph}}\right)^{2/3}.
  \label{eq-pbrNR-1D}
 \end{eqnarray}
In the ultrarelativistic case, when $\gamma_{ph}=1/\sqrt{1-\beta^2_{ph}}\gg1$, the $x-$component of the electron momentum is
$$p_x=  \qquad \qquad \qquad \qquad \qquad \qquad \qquad \qquad $$
\begin{equation}
p_{br}-m_ec\gamma_{br}^2\left(\frac{9}{2}\beta_{ph}\right)^{1/3}\left(k_{UH}\delta X\right)^{2/3}.
  \label{eq-pbr-1D}
 \end{equation}
 Here 
 \begin{equation}
 {k_{UH}=\frac{1}{c}\sqrt{\omega^2_{pe}+\frac{\omega^2_{Be}}{\gamma_{ph}}}}. 
  \label{eq-kUH}
 \end{equation}
 The value of 
 $2\pi/k_{UH}$ is an order of magnitude within the relativistic nonlinear upper hybrid wave length.
 
 In Fig. \ref{fig3} we show the profile of the breaking upper hybrid wave obtained from numerical integration of Eqs. (\ref{eq-px1DX} - \ref{eq-Ex1DX})
 for the phase velocity of the wave equal to $c\beta_{ph}=0.95 c$ and a magnetic field corresponding to $\omega_{Be}/\omega_{pe}=-2$. 
  It is clearly
 seen that at the wave breaking the wave profile becomes singular with the gradients of all the functions approaching infinity. 
 The y-component of the electron momentum changes direction at $X=X_{br}$. The electron density according to Eq. (\ref{eq-n1DX})
 tends to infinity as $n\propto \delta X^{-2/3}$.

\begin{figure}[tbph]
\centering
\includegraphics[width=6cm,height=4cm]{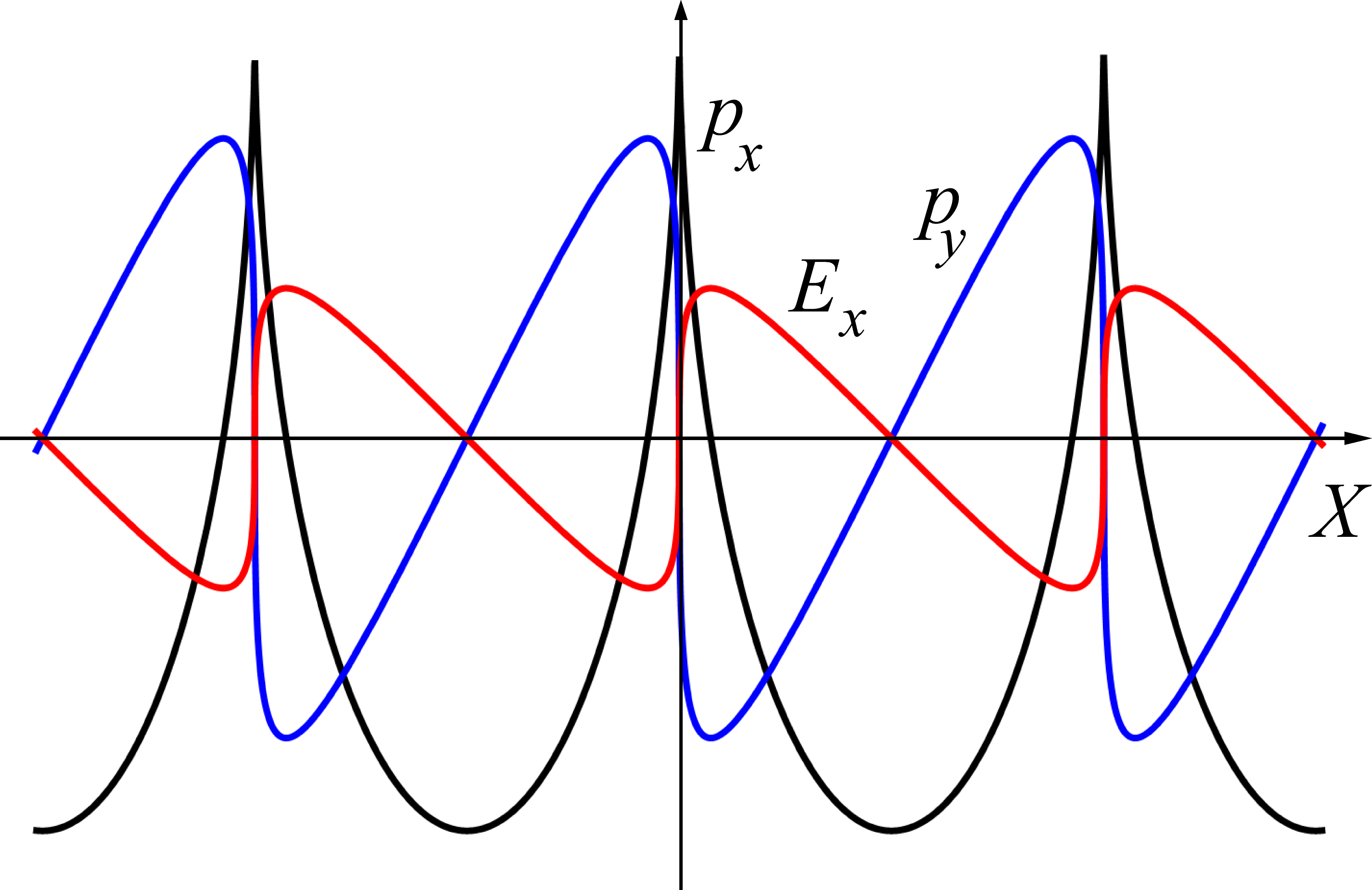}
\caption{Profile of the breaking upper hybrid wave for $\omega_{Be}/\omega_{pe}=-2$ and $\beta_{ph}=0.95$.}
\label{fig3}
\end{figure}

\section{Cylindrical Upper Hybrid Wave}

\subsection{The Lagrangian form of the equations describing cylindrical upper hybrid oscillations}

Here we consider axisymmetric cylindrical oscillations of the 
electron component of the plasma assuming the ions with homogeneous density $n_0$ to be motionless, 
the homogeneous constant magnetic field is directed along the z-axis, 
${\bf B}=B_0 {\bf e}_z$,
 and all the functions depend only on time $t$ and the spatial coordinate $r$.
In this case from the equations of the electron hydrodynamics and Maxwell's equations we obtain the following system:
 \begin{equation}
\partial_t n+\frac{1}{r}\partial_r (r n v_r)=0,
 \label{eq-Cont}
 \end{equation}
  \begin{equation}
\partial_t p_r+v_r\partial_r p_r-\frac{p_{\varphi}v_{\varphi}}{r}=-e E_r-\frac{e}{c} v_{\varphi} B_0,
 \label{eq-pr}
 \end{equation}
 \begin{equation}
\partial_t p_{\varphi}+v_r\partial_r p_{\varphi}+\frac{p_{\varphi}v_{r}}{r}=\frac{e}{c} v_r B_0,
 \label{eq-pphi}
 \end{equation}
  \begin{equation}
\partial_t E_r+v_r\frac{1}{r}\partial_r (r E_r)=4\pi n_0 e v_r.
 \label{eq-Er}
 \end{equation}
 Here $n$ is the electron density, $p_r$ and $p_{\varphi}$ are the radial and azimuthal components of 
 the electron momentum, and $E_r$ is the radial component of the electric field. The radial and azimuthal components of the 
electron velocity are equal to $v_{r}=c p_r/\sqrt{m_e^2c^2+p_r^2+p_{\varphi}^2}$ and $v_{\varphi}=c p_{\varphi}/\sqrt{m_e^2c^2+p_r^2+p_{\varphi}^2}$, respectively.

One of the most distinct difference between planar and cylindrical geometry is 
the appearance of the centrifugal force in the l.h.s. of Eq. (\ref{eq-pr}). 
The azimuthal momentum 
changes due to interaction of the radial motion with the magnetic field and vice versa. 
The effect of the centrifugal force becomes stronger the closer the radial 
flow of the electron component converges  thus preventing the flow collapse on the axis.

In order to solve Eqs. (\ref{eq-pr} - \ref{eq-Er}) it is convenient to change from 
 Euler coordinates, $r,t$, to Lagrange coordinates, $r_0,t$, \cite{DGBSS}. 
The Euler and Lagrange variables are related to each other as 
  \begin{equation}
r=r_0+\rho(r_0,t),
 \label{eq-EuLa}
 \end{equation}
where $\rho(r_0,t)$ is a displacement of the electron fluid elements from its initial position, i.e. $\rho(r_0,0)=0$.

In the Lagrange coordinates the continuity equation (\ref{eq-Cont}) has a solution 
  \begin{equation}
n(r_0,t)=\frac{n_0}{|J(r_0,t)|},
 \label{eq-EuLa}
 \end{equation}
where $J(r_0,t)=(r_0+\rho)(1+\partial_{r_0}\rho)/r_0$ is the Jacobian of the transformation from the Euler to Lagrange variables.
Equations (\ref{eq-pr} - \ref{eq-Er}) can be written in the form 
  \begin{equation}
\dot p_r-\frac{p_{\varphi}v_{\varphi}}{r}=-e E_r-\frac{e}{c} v_{\varphi} B_0,
 \label{eq-prLa}
 \end{equation}
 \begin{equation}
\dot p_{\varphi}+\frac{p_{\varphi} v_r}{r}=\frac{e}{c} v_r B_0,
 \label{eq-pphiLa}
 \end{equation}
  \begin{equation}
\dot E_r+\frac{E_r v_r}{r}=4\pi n_0 e v_r.
 \label{eq-ErLa}
 \end{equation}
 
 Taking into account that the radial and azimuthal components of the electron velocity are equal to 
 $v_r=\dot \rho$ and $v_{\varphi}=r \dot \varphi$ 
 with $\varphi$ being the azimuthal angle, we find solutions to Eq. (\ref{eq-pphiLa}) for the azimuthal momentum
 \begin{equation}
r p_{\varphi}=\frac{e}{2c} B_0 r^2+{\mathbb M}_{\varphi}(r_0)
 \label{eq-pphiLaM}
 \end{equation}
 and to Eq. (\ref{eq-ErLa}) for the electric field
 \begin{equation}
r E_r=2\pi n_0 e r^2 +{\mathbb Q}(r_0).
 \label{eq-ErLaQ}
 \end{equation}
 These expressions are the consequences of the conservation of the azimuthal component of the canonical momentum 
 and of the charge conservation, respectively. Here functions ${\mathbb M}_{\varphi}(r_0)$ and ${\mathbb Q}(r_0)$ are determined by
 the initial conditions. 
 
 Similar to the planar geometry case, when the electric field and the $y$-component of the electron momentum are related
 to each other via Eq. (\ref{eq-EB0}), in cylindrical geometry we have
 \begin{equation}
r \left(\frac{E_r}{4\pi n_0 e}-\frac{p_{\varphi} c}{e B_0}\right)=\frac{{\mathbb Q}(r_0)}{4\pi n_0 e}-\frac{{\mathbb M}_{\varphi}(r_0) c}{eB_0}.
 \label{eq-EB0MQ}
 \end{equation}

 Equations (\ref{eq-prLa}) and (\ref{eq-pphiLa}) with the electric field given by Eq. (\ref{eq-ErLaQ}) can be written in the Lagrangian form
 with the Lagrange function
 \[{\cal L}= -m_ec^2\sqrt{1-\frac{\dot r^2+r^2\dot \phi^2}{c^2}}\]
 \begin{equation}
+\frac{e B_0}{2 c}r^2\dot \phi-\pi n_0 e^2 r^2 +e{\mathbb Q}(r_0)\ln\frac{r}{r_0}.
 \label{eq-LagrFun}
 \end{equation}
 
 Since the Lagrange function (\ref{eq-LagrFun}) does not depend on time explicitly, the conservation of the Jacobi integral,
  \begin{equation}
  \frac{\partial {\cal L}}{\partial_{\dot q_j}}{\dot q_j}-{\cal L}=g(r_0),
   \label{eq-JacInt}
 \end{equation}
 where $q_j=(r,\varphi)$ and $\dot q_j=(\dot r,\dot \varphi)$, 
results in the energy conservation
 \begin{equation}
{\cal E}=-\pi n_0 e^2 r^2 +{\mathbb Q}(r_0) \ln \frac{r}{r_0} +g(r_0).
  \label{eq-Energy}
 \end{equation}
On the l.h.s. of this equation the energy is equal to 
 \begin{equation}
 {\cal E}=\sqrt{m_e^2c^4+p_{\varphi}^2+p_r^2}.
  \end{equation}
In the r.h.s. the function $g(r_0)$ is determined by the initial conditions.
 Using expressions (\ref{eq-pphiLaM}) and (\ref{eq-Energy}) it is easy to find the radial component of the electron momentum
  $p_r$ as a function of the coordinate $r$. Then with the equation for $v_r$, which is 
  \begin{equation}
\dot {r}=c^2\frac{p_r}{{\cal E}},
 \label{eq-radt}
 \end{equation}
 we solve by quadratures  Eqs. (\ref{eq-prLa} -\ref{eq-ErLa}), because the r.h.s. of Eq. (\ref{eq-radt})
 explicitly depends on the coordinate $r$ only. Its dependence on the Lagrange coordinate is parametric. For a known time dependence  of the 
 radial coordinate, $r(t)$, integration on time of the equation for the azimuthal angle
  \begin{equation}
\dot {\varphi}=c^2\frac{p_{\varphi}}{r{\cal E}}
 \label{eq-phit}
 \end{equation}
 yields the trajectory of the element of the electron fluid. Its projection to the $r, \varphi$ plane is given by
 \begin{equation}
\varphi-\int{\frac{p_{\varphi}dr}{r p_r}}={\rm constant}.
 \label{eq-phir}
 \end{equation}
 
 In the case when the initial value of the electric field and azimuthal momentum vanish, the functions ${\mathbb Q}(r_0)$ and ${\mathbb M}_{\varphi}(r_0)$ are  
 equal to ${\mathbb Q}(r_0)=-2\pi n_0 e r_0^2$ and ${\mathbb M}_{\varphi}(r_0)=-(e B_0/2 c) r^2_0$, respectively. If the radial momentum initial value equals 
 $p_{r,m}$, from Eq. (\ref{eq-Energy}) it follows
 \[
 \sqrt{ m_e^2c^4+\left(\frac{ e B_0}{2}\right)^2 \left(r-\frac{r_0^2}{r}  \right)^2 +p_r^2c^2
 }
 \]
 \begin{equation}
 =\sqrt{m_e^2c^4+p_{r,m}^2c^2}
 \label{eq-lim}
 \end{equation}
 \[-\pi n_0 e^2 \left(r^2-r_0^2\ln{\frac{r}{r_0}}\right).
 \]
 
 Fig. \ref{fig4} shows the energy iso-contours plotted for expression (\ref{eq-lim}) with $\omega_{Be}=2$ and $\omega_{pe}=1$
 \begin{figure}[tbph]
 \centering
\includegraphics[width=6cm,height=6cm]{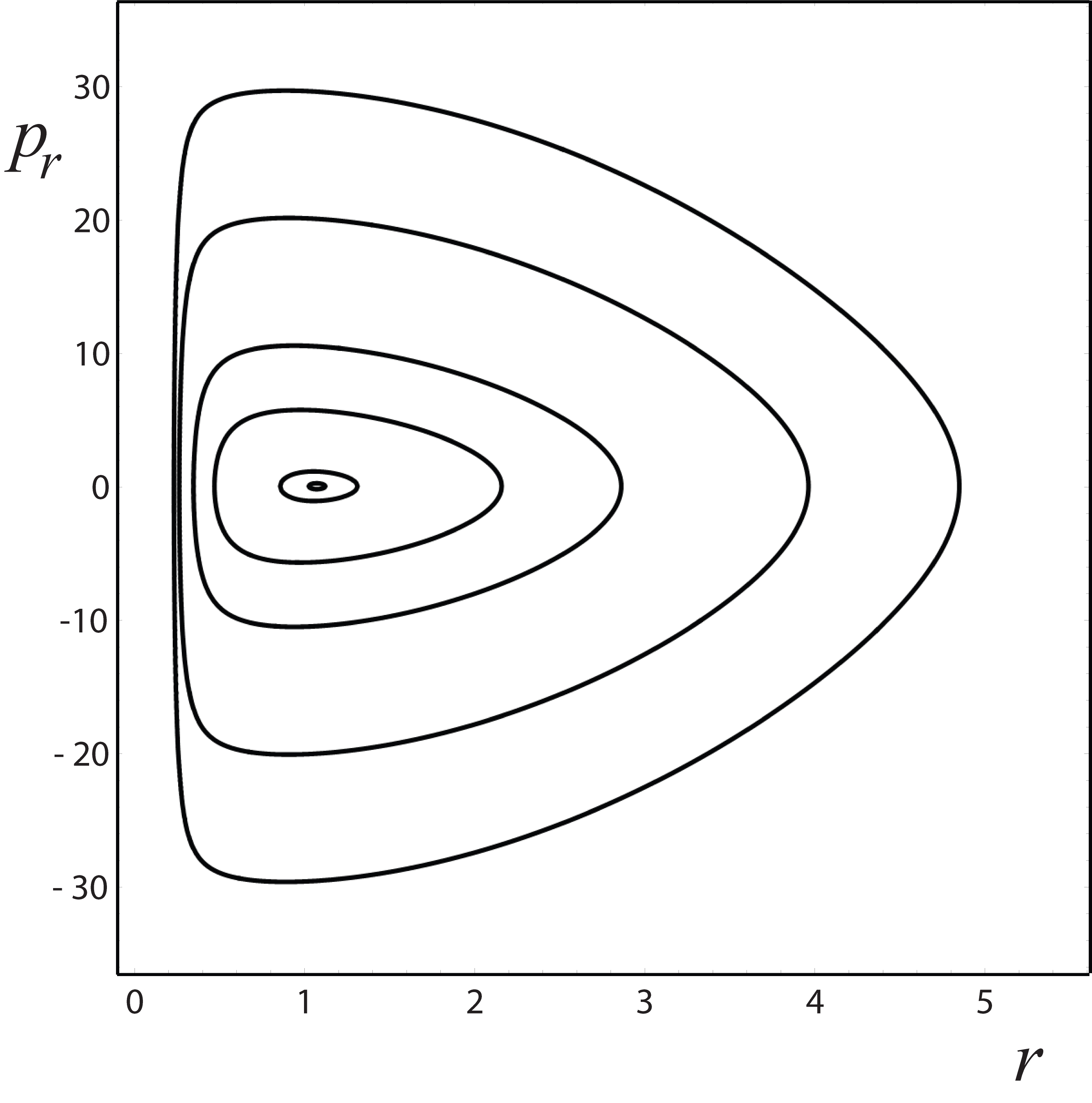}
\caption{Iso-energy contours for expression (\ref{eq-lim}) for $\omega_{Be}=2$ and $\omega_{pe}=1$ and $h=9,18,27,36,45$.}
\label{fig4}
\end{figure}
As we see the radial coordinate cannot vanish due the logarithmic divergence of the electrostatic potential and the $\propto 1/r$ divergence 
of the ``centrifugal energy" at $r\to 0$. The minimal value of the radius can be estimated to be approximately 
equal to
\begin{equation}
r_{\min}=r_0^2\frac{eB_0}{p_{r,m}c}.
 \label{eq-rmin}
 \end{equation}
 
 Fig. \ref{fig5} presents the typical electron trajectory in the $(x=r\cos{\varphi},y=r\sin{\varphi})$ 
 plane for $r_0=1.5$, $p_r(0)=-0.5$, $p_{\varphi}(0)=-0.25$, and $\omega_{Be}/\omega_{pe}=0.25$. 
 It is clearly seen that the trajectory is localized in the region $r>r_{\min}$,
 which, in the case under consideration, is about 0.31.
 \begin{figure}[tbph]
 \centering
\includegraphics[width=5cm,height=5cm]{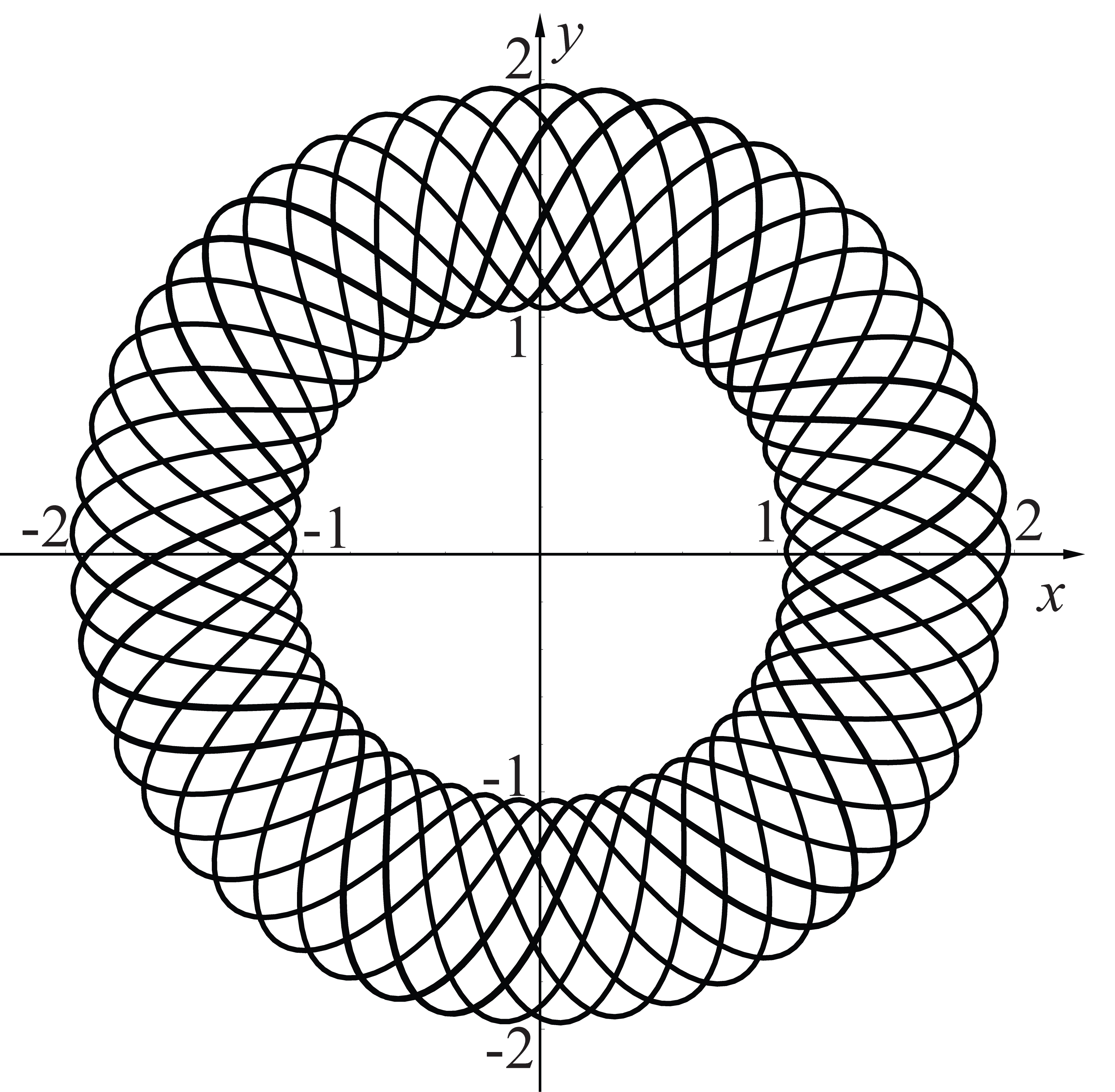}
\caption{Electron trajectory in the $(x=r\cos{\varphi},y=r\sin{\varphi})$ 
 plane for $r_0=1.5$, $p_r(0)=-0.5$, $p_{\varphi}(0)=-0.25$, and $\omega_{Be}/\omega_{pe}=0.25$.}
\label{fig5}
\end{figure}
 
  \subsection{On the cavity closing in a wake left behind the laser pulse}
  As is well known, the cavity, the positively charged region void of electrons, 
  is formed because almost all the electrons in the way of
a sufficiently intense laser pulse are pushed aside \cite{PMtV, BW2}.  The unperturbed electrons on the side are attracted 
by the positive charge. They close the cavity at the finite distance behind the laser pulse.
A non-zero axial magnetic field prevents the cavity from closing as shown in the previous section. 
The process of cavity closing is explained in Fig. \ref{fig6} in the case of 
vanishing magnetic field (Fig. \ref{fig6} a) and for finite magnetic field (Fig. \ref{fig6} b) when it results 
in the appearance of a hole at the rear of the cavity.
\begin{figure}
\centering
\includegraphics[width=6cm,height=3cm]{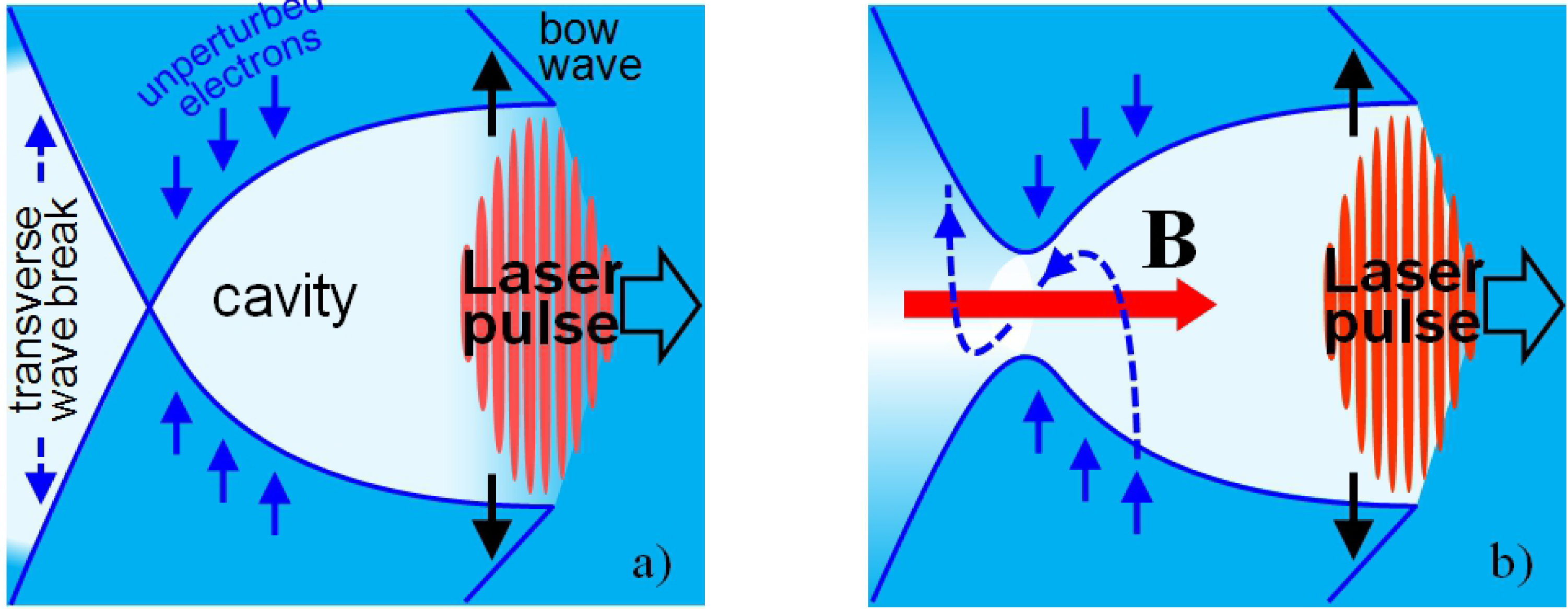}
\caption{The cavity formed behind the laser pulse in a collisionless plasma.
The walls formed by unperturbed electrons converge at the rear of the cavity (a).
In the presence of a longitudinal magnetic field, the electrons miss the axis, 
which results in the appearance of a hole at the rear of the cavity (b).} 
\label{fig6} 
\end{figure}

In order to estimate the radius of the hole at the rear wall of the cavity we choose the parameters of the converging cylindrical 
wave as follows. The functions ${\mathbb M}_{\varphi}(r_0)$ and ${\mathbb Q}(r_0)$ determining the azimuthal momentum in Eq. (\ref{eq-pphiLaM})
and the radial component of the electric field in Eq. (\ref{eq-ErLaQ}), respectively, are assumed to be equal to ${\mathbb M}_{\varphi}(r_0)=0$
and
\begin{equation}
{\mathbb Q}(r_0)= \left \{
	\begin{array}{lcr}
		 0, & r_0 \leq r_c; \\
		 2\pi n_0 e(r_c^2-r_0^2), & r_0 > r_c. \\
	\end{array}
	\right.
\label{eq-Qr0rm}
\end{equation} 
Here $r_c$ is the initial cavity radius.
Using this expression we obtain that at $t=0$ the electric field is a linear function of radius for $r_0 \leq r_c$ and it 
is inversely proportional to $r_0$ at $r_0 > r_c$ (see Fig. \ref{fig7}),
\begin{equation}
E_r(r_0)= - 2\pi n_0 e\left \{
	\begin{array}{lcr}
		 r_0, & r_0 \leq r_c; \\
		 r_c^2/r_0, & r_0 > r_c, \\
	\end{array}
	\right.
\label{eq-Ercr0}
\end{equation} 
in accordance with the radial electric field profile found with the 3D PIC simulations (see Fig. 2 d,e in Ref. \cite{BW2}). 
\begin{figure}[tbph]
\centering
\includegraphics[width=6cm,height=3.5cm]{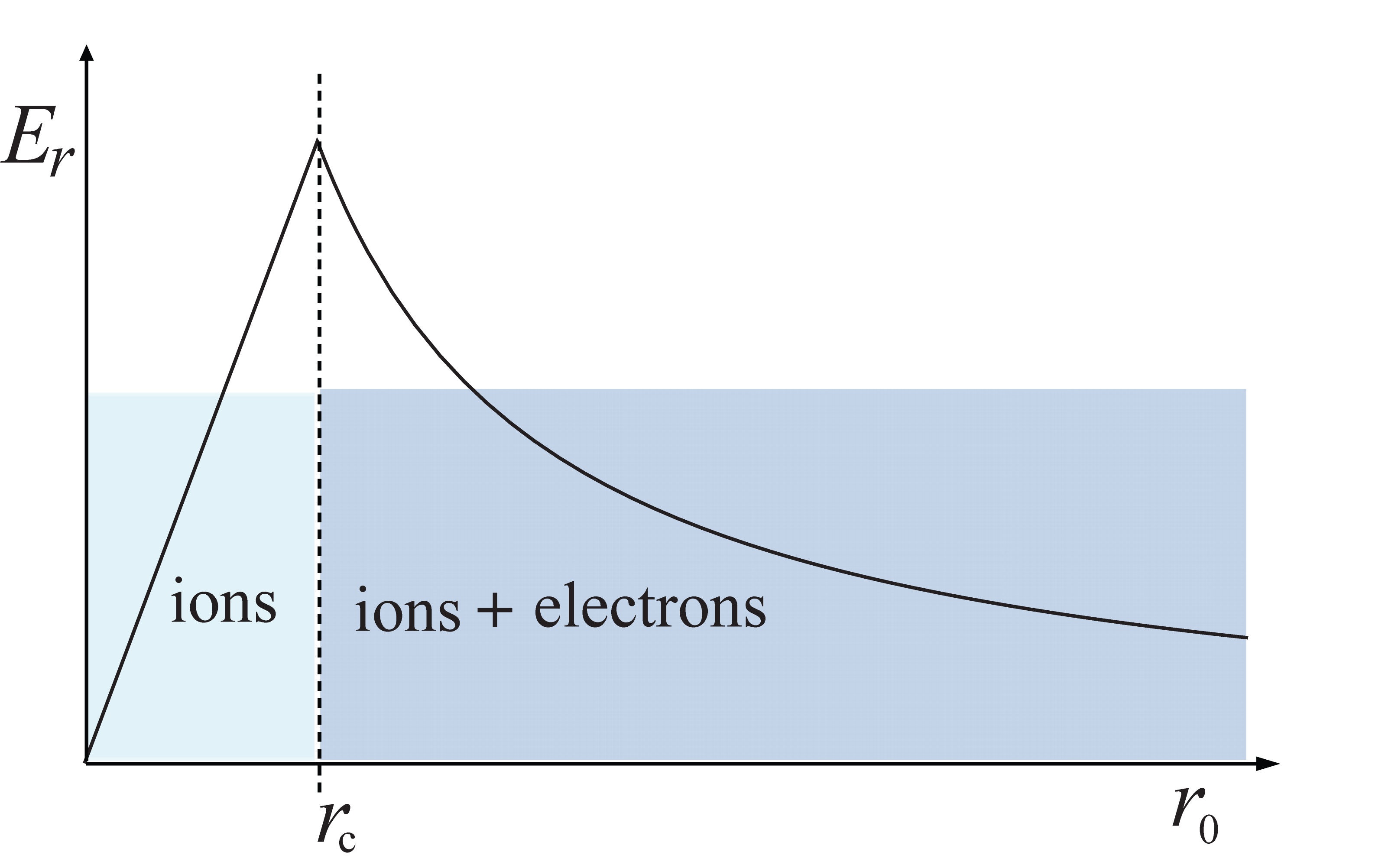}
\caption{Electric field as a function of the Lagrange coordinate, $r_0$.}
\label{fig7}
\end{figure}
Considering motion of the electron fluid element with $r_0=r_c$ 
and the initial value of the radial momentum  equal to zero we obtain from Eq. (\ref{eq-Energy})
\[\sqrt{1+\left(\frac{ \omega_{Be}^2}{2  c^2}\right)
 \left( r-\frac{r_c^2}{r} \right)^2
 +\left(\frac{p_r}{m_e c}\right)^2 }-1
\]
 \begin{equation}
 =\frac{\omega_{pe}^2 }{4c^2}\left(r_c^2-r^2\right).
 \label{eq-closing}
 \end{equation}
According to Eq. (\ref{eq-closing}) the radial component of the electron momentum vanishes at $r=r_c$ and at $r=r_{\min}$, which is given by equation 
\[ \sqrt{1+\left(\frac{\omega_{Be}^2}{2 c^2}\right)
 \left( r_{\min}-\frac{r_c^2}{r_{\min}} \right)^2}-1\]
 \begin{equation}
=\frac{\omega_{pe}^2 }{4c^2}\left(r_c^2-r_{\min}^2\right).
 \label{eq-r_min}
 \end{equation}
 
 In Fig. \ref{fig8} we plot dependence on the magnetic field ($\omega_{Be}/\omega_{pe}$) of the minimum radius, $r_{\min}$, 
 normalized by $2 d_e=c/\omega_{pe}$, for different values of
 the cavity radius, $r_c$.
 \begin{figure}[tbph]
 \centering
\includegraphics[width=5cm,height=5cm]{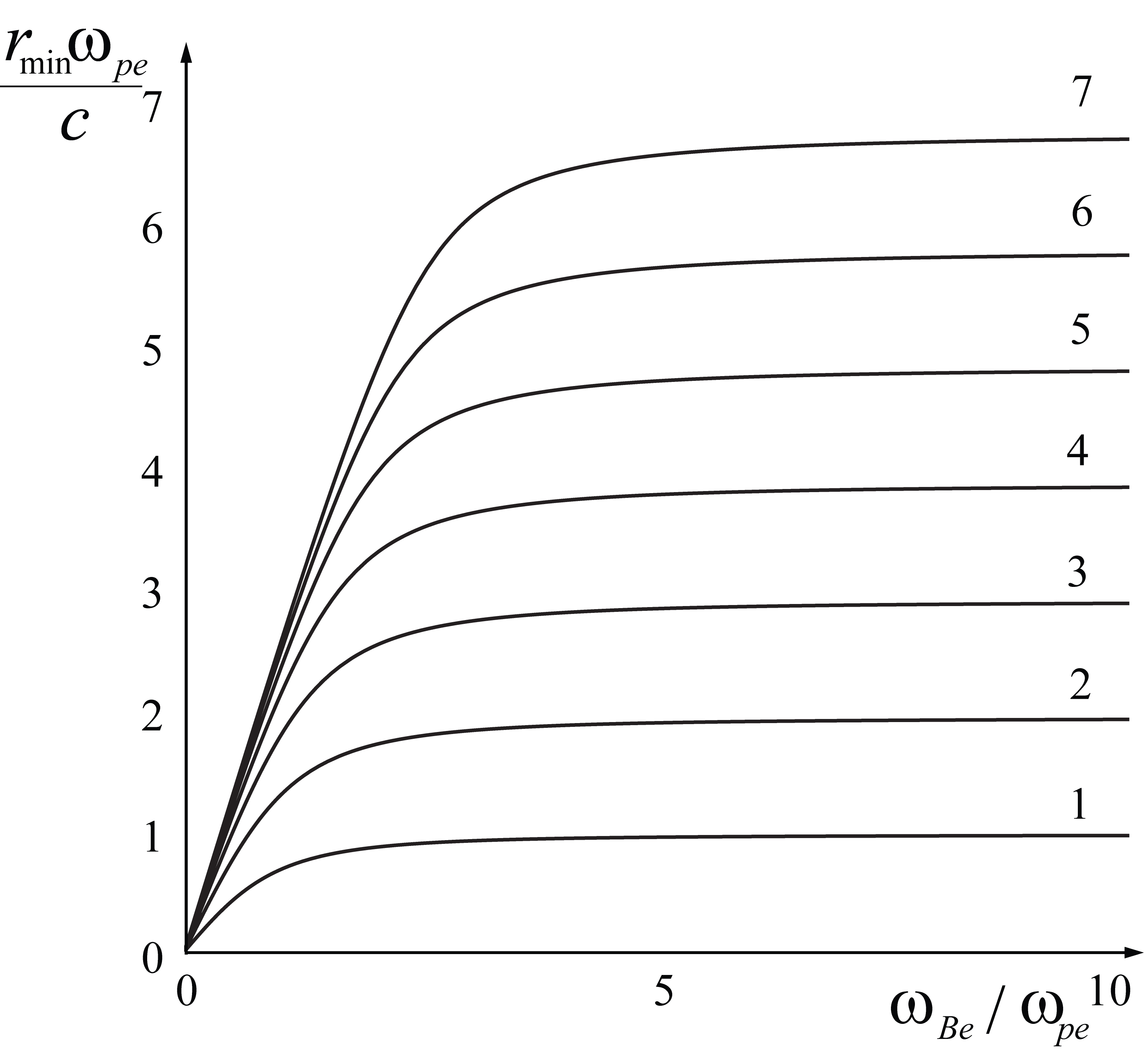}
\caption{Minimum radius, $r_{\min}$, normalized by $d_e=c/\omega_{pe}$ v.s. the parameter (\ref{eq-epsB}) characterizing the magnetic field amplitude, 
for the cavity radius, $r_c=1,2,3,4,5,6,7$}
\label{fig8}
\end{figure}
In the limit of weak magnetic field, when $\omega_{Be}/\omega_{pe}\ll 1$, the minimal radius, $r_{\min}$ 
is linearly proportional to $\omega_{Be}$, being approximately equal to
 \begin{equation}
 r_{\min}\approx \frac{4}{\sqrt{2}}\frac{c \omega_{Be}}{\omega^2_{pe}},
 \label{eq-r_minBweak}
 \end{equation}
 i.e. it does not depend on the initial cavity radius, $r_c$. For $B\to 0$ the hole radius tends to zero.
 In the limit of strong magnetic field, $\omega_{Be}/\omega_{pe}\gg 1$, we have $r_{min}\approx r_c$, 
 i.e. it weakly depends on the magnetic field amplitude.
 
 Fig. \ref{fig9} presents the typical electron trajectory in the $(x=r\cos{\varphi},y=r\sin{\varphi})$ 
 plane for $r_m=r_0=1$, $p_r(0)=0$, $p_{\varphi}(0)=0$, and $\omega_{Be}/\omega_{pe}=0.5$. 
 It is clearly seen that the trajectory is localized in the region $r>r_{\min}$.
 \begin{figure}[tbph]
 \centering
\includegraphics[width=5cm,height=5cm]{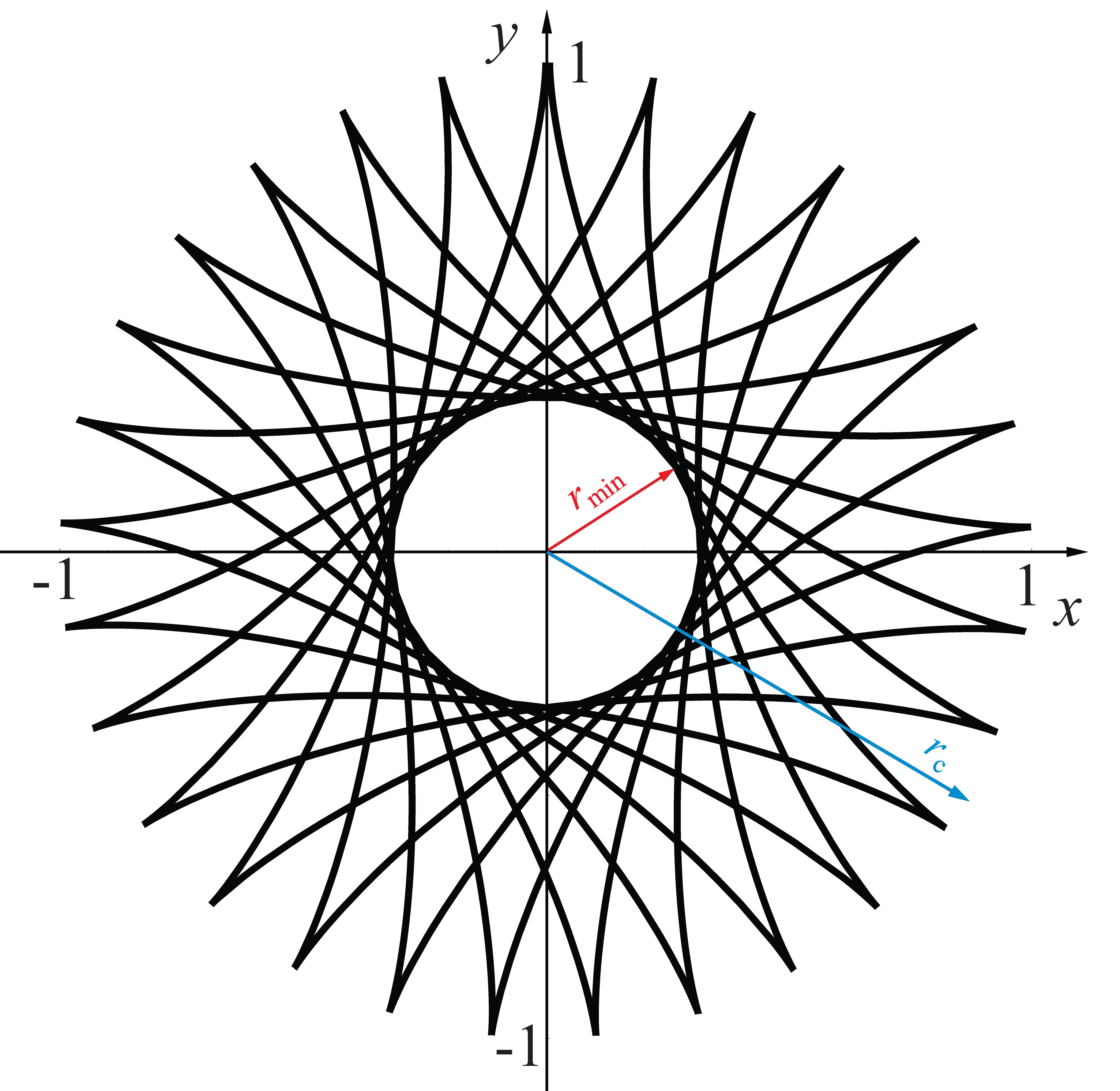}
\caption{Electron trajectory in the $(x=r\cos{\varphi},y=r\sin{\varphi})$ 
 plane for $r_m=r_0=1$, $p_r(0)=0$, $p_{\varphi}(0)=0$, and $\omega_{Be}/\omega_{pe}=0.5$.}
\label{fig9}
\end{figure}

By using values of the Langmuir and Larmor frequencies, $\omega_{pe}=5.642\times 10^4 \sqrt{n_0(cm^{-3})} s^{-1}$ and $\omega_{Be}=1.759\times 10^11 B(T) s^{-1}$,
we estimate the minimal radius for $n_0=10^{18} cm^{-3}$ and $B=10T$ to be $r_{\min}=4.677\times 10^{-5} cm$.

  \subsection{Betratron oscillations of relativistic electrons inside the wake wave cavity in the presence of the axial magnetic field}

The effects of axial magnetic field on the LWFA has been addressed in Refs. \cite{Schmit2012} and \cite{Bourdier2012} 
where the plasma densification through magnetic
compression has been studied in order to overcome dephasing between the wake field and accelerated electrons and the magnetic field effects 
on the electron injection have been analysed, respectively. Here we pay attention 
to the electron trajectory rotation in the azimuthal direction resulting in the azimuthal patterns of the LWFA accelerated electron bunches
which has been observed in the experiments reported in Ref. \cite{THBf}. The betratron oscillations in the laser wake field themselves provide 
a promising source of ultra short x-ray beam generation \cite{BetaX}. The magnetic field can be used for additional manipulation of the betatron oscillations 
and for controlling the properties of the radiation emitted. 

Inside the cavity formed in a plasma in a wake behind an ultra short laser pulse the electric field has a radial component $E_r=2\pi n e r$. 
Following betatron oscillation theory \cite{BetaTheory1} we assume that the longitudinal (axial) component of the electron momentum, $p_x$, is given.
Choosing as a function of time
 \begin{equation}
 p_x(t)=p_m (t_{acc}^2-t^2)/t_{acc}^2
 \label{eq-pxt}
 \end{equation}
 we describe the electron acceleration in the wake field  \cite{BetaTheory2}. During the acceleration time $t_{acc}\approx (\pi/\omega_{pe})\gamma_{ph}^2$ the 
 electron longitudinal momentum changes from zero to $p_m\approx m_e c \gamma_{ph}^2$ \cite{ESW}. 
 Here the gamma factor associated with the phase velocity of the wake wave is equal to $\gamma_{ph}=\sqrt{\omega_{0}/\omega_{pe}}$. The equations 
 of transverse electron motion are identical to Eqs. (\ref{eq-prLa}) and (\ref{eq-pphiLa}) where the radial electric field is given by $E_r=2\pi n e r$. 
 The electron velocity is 
 ${\bf v}=(p_x/\gamma, p_r/\gamma, p_{\varphi}/\gamma)$ with the relativistic gamma factor $\gamma=\sqrt{1+p_x^2+p_r^2+p_{\varphi}^2}$.

 In the case when the magnetic field vanishes, the electron trajectory is described by a linear combination of associated Legendre functions \cite{BetaTheory2}.
 If the electron is injected into the wake wave with zero azimuthal momentum its trajectory lies in the same plane. The axial magnetic field causes the electron 
 trajectory to rotate as seen in Fig. \ref{fig10}. In this figure we plot the electron trajectory obtained via numerical integration of the electron equations 
 of motion for a normalized magnetic field $B=0.25$, initial coordinates, $r_0=1$, $\varphi_0=\pi/2$, $x_0=-800$, initial momentum $p_x=0$, $p_r=0$, $p_{\varphi=0}$,
 acceleration time $t_{acc}=800$ and maximal longitudinal momentum equal to to $p_m=50$, respectively.
 \begin{figure}[tbph]
 \centering
\includegraphics[width=6cm,height=5.5cm]{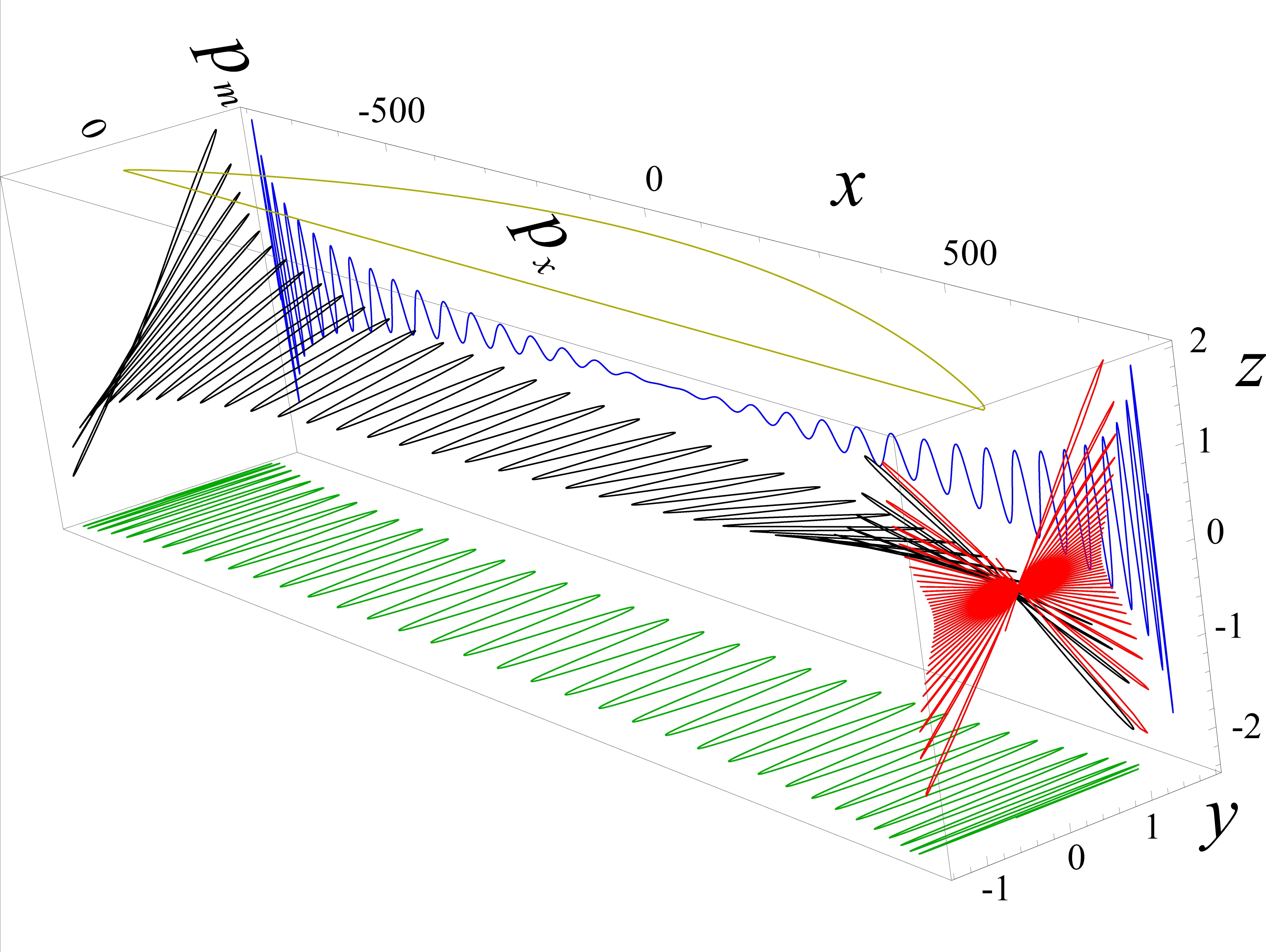}
\caption{Betatron oscillations for $B=0.25$, $r_0=1$, $\varphi_0=\pi/2$, $x_0=-800$, $p_x=0$, $p_r=0$, $p_{\varphi=0}$, $t_{acc}=800$, and $p_m=50$.
The electron trajectory (black curve) is confined in a slowly rotating plane. The trajectory 
 projections on the bottom of the $x,y$ plane(green curve) and on the right $x,z$ plane (blue curve) 
 show slowly varying amplitudes and periods of the betaron oscillations. In the upper $x,y$ 
 plane we plot the longitudinal momentum $p_x$ as a function of the $x$-coordinate (yellow).
 The trajectory projection on the r.h.s. $y,z$ plane (red) shows a "star"-like pattern. }
\label{fig10}
\end{figure}
 \begin{figure}[tbph]
 \centering
\includegraphics[width=6cm,height=3cm]{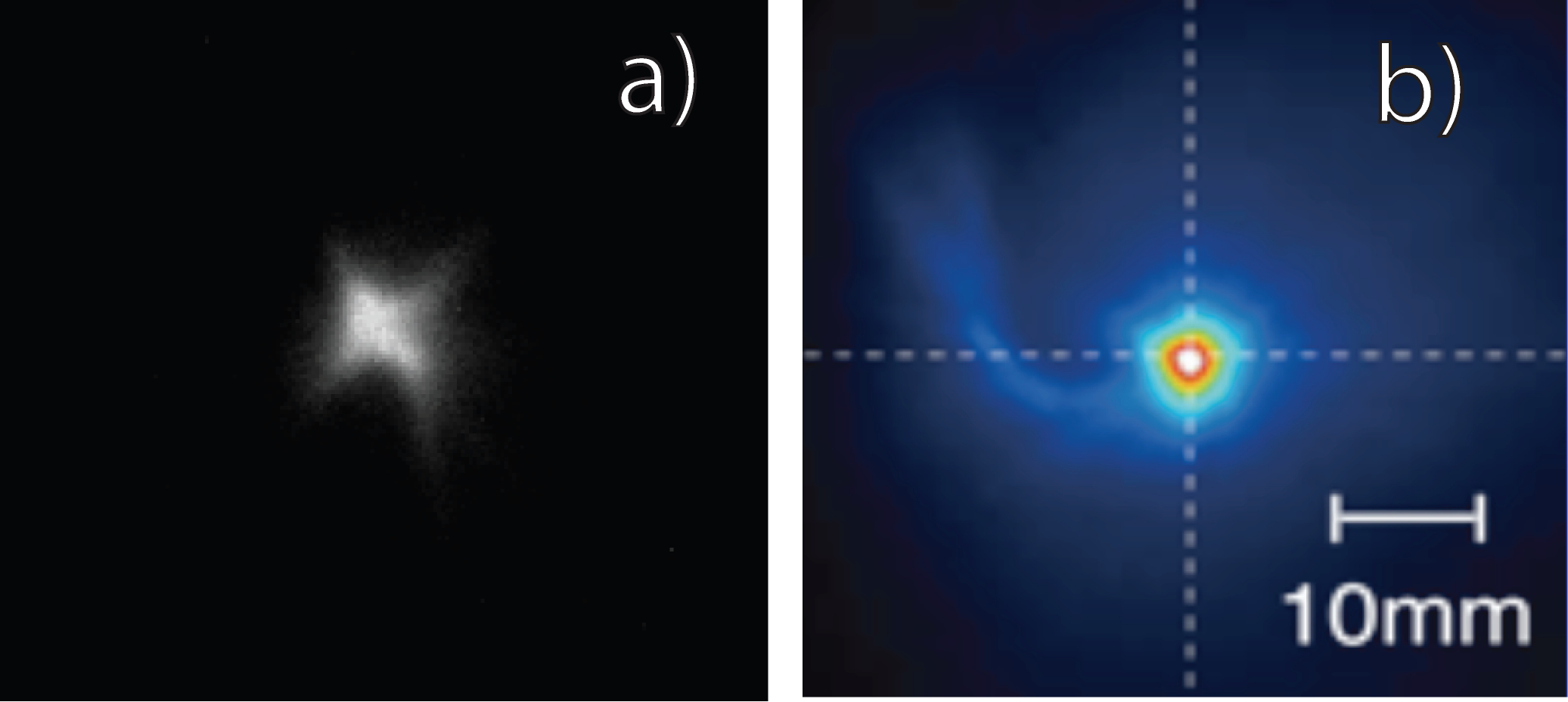}
\caption{ Images of the electron beam generated with $B =0.2$T, gas density $n_{He}=4 \times 10^{19} {\rm cm}^3$, and a laser power of 12 TW. 
The laser polarization
is in the vertical direction.
a) Four-ray star. b) Spiral pattern.}
\label{fig11}
\end{figure}

 We distinctly see the betatron oscillations of the electron whose amplitude and period changes according to dependence of the betatron frequency on the 
 electron energy, $\omega_b=\omega_{pe}/\gamma$, and by virtue of the adiabatic invariant conservation ${\cal E}_{perp}/\omega_b={\rm constant}$ \cite{BetaTheory2}.
 Here ${\cal E}_{perp}$ is the kinetic energy of the transverse motion. The electron trajectory (black curve) is confined 
 on a slowly rotating plane. The trajectory 
 projections on the bottom of the $x,y$ plane(green curve) and in the right $x,z$ plane (blue curve) 
 show slowly varying amplitudes and the period of the betaron oscillations. On the upper $x,y$ 
 plane we plot the longitudinal momentum $p_x$ as a function of the $x$-coordinate (yellow).
 The trajectory projection on the r.h.s. $y,z$ plane (red) shows a ``star"-like pattern. 
 We note an analogy of this pattern with the images of the electron beam generated in the experiments \cite{THBf} on LWFA with an axial magnetic field.
  The ``star-like" pattern should also be revealed in the betatron emission. In Fig. \ref{fig11} we present images of the LWFA electron beam observed  in 
  \cite{THBf} for the axial magnetic field equal to $0.2$ T. We see a ``Four-Ray Star" image in Fig. \ref{fig11} a) and a spiral pattern in Fig. \ref{fig11} b),
  which clearly indicates the magnetic field effect.
 
 When the magnetic field vanishes a typical electron beam image has an elliptic form,
 as in Fig. \ref{fig12}. The electron trajectory (black curve) has the form of a spiral with a slowly changing amplitude and period. The trajectory 
 projections on the bottom of the $x,y$ plane(green curve) and on the right $x,z$ plane (blue curve) 
 show slowly varying amplitudes and the period of the betaron oscillations. On the upper $x,y$ 
 plane we plot the longitudinal momentum $p_x$ as a function of the $x$-coordinate (yellow).
 The trajectory projection on the r.h.s. $y,z$ plane (red) shows an ``ellipse"-like pattern. 
 Such elliptic structures have been discussed in detail and have been detected in experiments \cite{BetaImage}.
 \begin{figure}[tbph]
 \centering
\includegraphics[width=6cm,height=5.5cm]{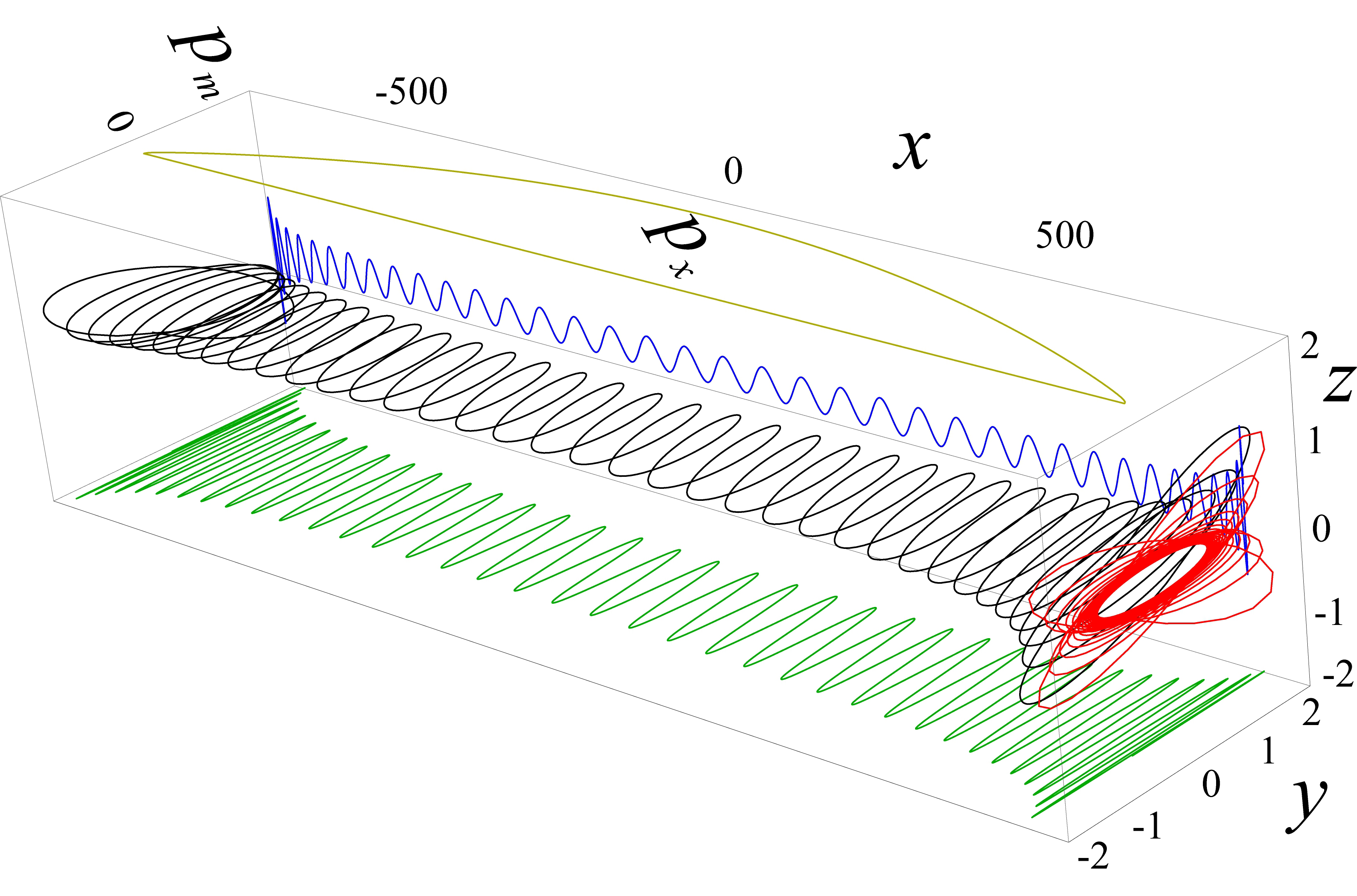}
\caption{Betatron oscillations for $B=0$, $r_0=1$, $\varphi_0=\pi/2$, $x_0=-800$, $p_x=0$, $p_r=0$, $p_{\varphi=1}$, $t_{acc}=800$, and $p_m=50$.
The electron trajectory (black curve) has the form of a spiral with slowly changing amplitude and period. The trajectory 
 projections on the bottom of the $x,y$ plane(green curve) and on the right $x,z$ plane (blue curve) 
 show slowly varying amplitudes and the period of the betaron oscillations. On the upper $x,y$ 
 plane we plot the longitudinal momentum $p_x$ as a function of the $x$-coordinate (yellow).
 The trajectory projection in the r.h.s. $y,z$ plane (red) shows an "ellipse"-like pattern.}
\label{fig12}
\end{figure}

\section{Computer Simulation Results}
We note here that the above analyzed properties of nonlinear plasma waves in magnetized plasmas cannot be revealed in a 2D planar $(x,y)$ symmetry configuration 
due to the vanishing of the centrifugal force.
In order to reveal the 3D effects in the configuration typical for the experiments in high power ultrashort pulse laser interaction 
with underdense plasmas we performed three dimensional Particle in Cell simulations with the code REMP \cite{REMP} whose results are presented in Fig. \ref{fig13}. The Gaussian linearly polarized 
(along $z$-axis) laser pulse with the amplitude of $a_0=6$ and FWHM dimensions of $8\lambda \times 20\lambda \times 20\lambda$ 
propagates in homogeneous plasma with the density of $n_e=0.0009n_{cr}$.
Ions are assumed to be immobile.
The simulation box size is $80\lambda \times 72\lambda \times 72\lambda$, 
the mesh sizes are $dx = \lambda/16$, $dy=dz=\lambda/8$, and the total number of quasi-particles is $6.3\times 10^8$.

Fig. \ref{fig13} (a,b,c) presents the electron density in the wake of the laser pulse propagating from left to right in a homogeneous plasma, when 
no static magnetic field is imposed.  
In Fig. \ref{fig13} (d,e,f) we plot the electron density distribution in the presence of a homogeneous permanent longitudinal magnetic field of 10 T.
Half of the box is removed to reveal the interior (a,d), 
a close-up of the isosurface corresponding to $n_e/n_{cr}=0.008$ 
is shown in the location of the wave-breaking (b,e); 
the cross section is about $x=60.9\lambda$ averaged over a half of the laser wavelength (c,f).
In Fig. \ref{fig13} (a,b,c) we see the cavity formed in the electron density with thin high density walls. The rear side of the cavity is 
closed resulting in the electron bunch injection along the axis. The electron density has a local maximum on the axis (Fig. \ref{fig13} (b,c)).
The presence of the static magnetic field prevents the cavity from  closing (Fig. \ref{fig13} (a,b)) in accordance with the above formulated 
theory of the cylindrical Upper Hybrid plasma wave. The electron bunch injected along the cavity axis from its rear side is ring-shaped with no electrons 
on the axis. The electron density cross section reveals the ``Four-Ray Star" pattern which has been observed in the experiments \cite{THBf}. 

In the region relatively far from the axis where the magnetic field is relatively small, Fig. \ref{fig13} (a,d), the electron density distribution  
is not affected by the magnetic field showing identical shapes of the cavity walls and bow wave.

 \begin{figure*}[tbph]
\includegraphics[width=12cm,height=6cm]{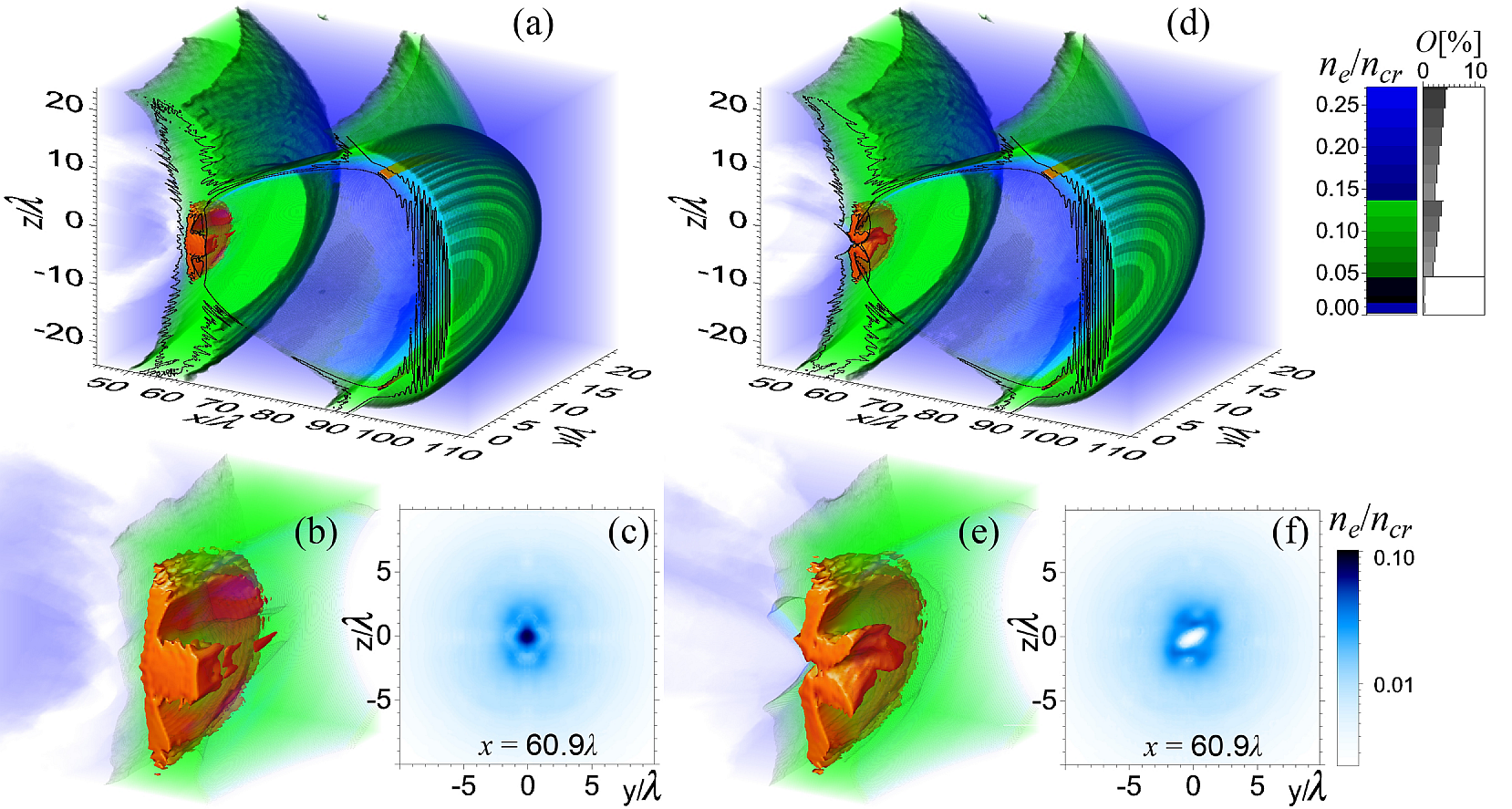}
\caption{Electron density in the wake of the laser pulse propagating from left to right in a homogeneous plasma 
with no static magnetic field initially imposed (a,b,c) 
and in the presence of a longitudinal magnetic field of 10 T (d,e,f).
Half of the box is removed to reveal the interior (a,d), 
a close-up of the isosurface corresponding to $n_e/n_{cr}=0.008$ 
is shown in the location of the wave-breaking (b,e); 
the cross section is about $x=60.9\lambda$ averaged over a half of the laser wavelength (c,f).
}
\label{fig13}
\end{figure*}
\section{Conclusion}

The effect of the axial magnetic field on the cylindrical electrostatic 
wave dynamics leads to the precession of the electron trajectory 
in the $(x=r\cos{\varphi},y=r\sin{\varphi})$ plane. 
Conservation of the generalized momentum leads to a non-zero 
angular momentum of the electron component. 
As a result the electrons do not reach the axis remaining confined 
in the region $r>r_{\min}$ thus preventing  the 
rear wall of the cavity  in the wake behind ultra-short laser pulse from closing. 
In the limit of weak magnetic field, when $\omega_{Be}/\omega_{pe}\ll 1$, 
the minimal radius, $r_{\min}$ is linearly proportional to $\omega_{Be}$, 
as given by Eq. (\ref{eq-r_minBweak}). The magnetic field causes a rotation 
of the fast electron trajectory, which is seen in a typical ``Four-Ray Star"  
pattern of the electron density image observed in the experiments \cite{THBf} (see also Fig. \ref{fig11} and in the 3D PIC simulations 
 in Fig. \ref{fig13}).



\begin{thebibliography}{99}

\bibitem{VLG} V. L. Ginzburg, {\it The Propagation of Electromagnetic Waves in Plasmas} 
(Pergamon, Oxford, 1970).

\bibitem{FFC} F. F. Chen, {\it Introduction to Plasma Physics and Controlled Fusion}
(Springer, Berlin, 2010).

\bibitem{MTB}G. A. Mourou, T. Tajima, S. V. Bulanov, 
Rev. Mod. Phys. \textbf{78}, 309 (2006).

\bibitem{TD}T. Tajima and J. M. Dawson, Phys. Rev. Lett. \textbf{43}, 267 (1979).

\bibitem{ESW} E. Esarey, C. B. Schroeder, W. P. Leemans,
Rev. Mod. Phys. {\bf 81}, 1229 (2009).

\bibitem{PC} P. Chen, J. M. Dawson, R. W. Huff {\it et al.}, Phys. Rev. Lett. {\bf 54}, 693 (1985);
T. Katsouleas, Phys. Rev. A {\bf 33}, 2056 (1986).

\bibitem{LWFA} W. P. Leemans, B. Nagler, A. J. Gonsalves, C. Toth, K. Nakamura, C. G.
R. Geddes, E. Esarey, C. B. Schroeder, and S. M. Hooker, Nat. Phys. {\bf 2}, 696 (2006); 
I. Blumenfeld, C. E. Clayton, F.-J. Decker {\it et al.}, Nature (London) {\bf 445}, 741 (2007);
S. Karsch, J. Osterhoff, A. Popp, T. P. Rowlands-Rees, Zs. Major, M.
Fuchs, B. Marx, R. Horlein, K. Schmid, L. Veisz, S. Becker, U. Schramm,
B. Hidding, G. Pretzler, D. Habs, F. Gruener, F. Krausz, and S. M. Hooker, New J. Phys. {\bf 9}, 415 (2007);
M. Hafz, T. M. Jeong, I. W. Choi, S. K. Lee, K. H. Pae, V. V. Kulagin, 
J. H. Sung, T. J. Yu, K.-H. Hong, T. Hosokai, J. R. Cary, D.-K. Ko, and J. Lee, Nat. Photonics {\bf 2}, 571 (2008);
C. E. Clayton, J. E. Ralph, F. Albert, R. A. Fonseca, S. H. Glenzer, C. Joshi, W. Lu, K. A. Marsh, S. F. Martins, 
W. B. Mori, A. Pak, F. S. Tsung, B. B. Pollock, J. S. Ross, L. O. Silva, and D. H. Froula, Phys. Rev. Lett. {\bf 105}, 105003 (2010).

\bibitem{HHG} D. F. Gordon, B. Hafizi, D. Kaganovich, and A. Ting, Phys. Rev. Lett. {\bf 101}, 045004 (2008);  U. Teubner and P. Gibbon, Rev. Mod. Phys. {\bf 81}, 445 (2009); 
A. S. Pirozhkov, M. Kando, T. Zh. Esirkepov {\it et al.}, Phys. Rev. Lett. {\bf 108}, 135004 (2012).

\bibitem{FM} S. V. Bulanov, T. Zh. Esirkepov, and T. Tajima, Phys. Rev. Lett. \textbf{91}, 085001 (2003);
S. S. Bulanov \textit{et al.}, Phys. Rev. E \textbf{73}, 036408 (2006);
 V. V. Kulagin \textit{et al.}, Phys. Plasmas \textbf{14}, 113101 (2007); 
 D. Habs \textit{et al.}, Appl. Phys.B \textbf{93}, 349 (2008);
A. G. Zhidkov {\it et al.,} Phys. Rev. Lett. \textbf{103}, 215003 (2009);
 H. Wu \textit{et al.}, ibid. \textbf{104}, 234801 (2010); L. L. Ji \textit{et al.},
ibid. \textbf{105}, 025001 (2010); S. S. Bulanov, \textit{et al.}, Phys. Lett. A \textbf{374}, 476 (2010);
M. Wen {\it et al.}, Appl. Phys. Lett. {\bf 101}, 021102 (2012); 
H.-C. Wu and J. Meyer-ter-Vehn, Nat. Photonics {\bf 6}, 304 (2012);
M. Kando \textit{et al}., Phys. Rev. Lett. \textbf{99}, 135001 (2007); 
A. S. Pirozhkov, {\it et al.}, Plasma Phys. \textbf{14}, 123106 (2007)
M. Kando \textit{et al.}, Phys. Rev. Lett. \textbf{103}, 235003 (2009);
J. K. Koga, S. V. Bulanov, T. Zh. Esirkepov, A. S. Pirozhkov, M. Kando, and N. N. Rosanov,
 Phys. Rev. A. {\bf 86}, 053823 (2012); 
M. Lobet \textit{et al.}, Phys. Lett. A {\bf 377}, 1114 (2013).

\bibitem{BUFN} S. V. Bulanov, T. Zh. Esirkepov, M. Kando, A. S. Pirozhkov, and N. N. Rosanov, Physics Uspekhi (2013) in press.

\bibitem{INJ} S. V. Bulanov, I. N. Inovenkov, V. I. Kirsanov {\it et al.}, Phys. Fluids B {\bf 4}, 1935 (1992); 
C. A. Coverdale, C. B. Darrow, C. D. Decker {\it et al.}, Phys. Rev. Lett. {\bf 74}, 4659 (1995); 
A. Modena, A. Najmudin, E. Dangor {\it et al.}, Nature (London) {\bf 377}, 606 (1995); 
 D. Gordon, K. C. Tzeng, C. E. Clayton {\it et al.}, Phys. Rev. Lett. {\bf 80}, 2133 (1998); 
 S. V. Bulanov, N. Naumova, F. Pegoraro, and J. Sakai, Phys. Rev. E {\bf 58}, R5257 (1998); 
 H. Suk, N. Barov, J. B. Rosenzweig, and E. Esarey, Phys. Rev. Lett. {\bf 86}, 1011 (2001); 
M. C. Thompson, J. B. Rosenzweig, and H. Suk, Phys. Rev. ST Accel. Beams {\bf 7}, 011301 (2004); 
P. Tomassini, M. Galimberti, A. Giulietti {\it et al.}, Laser Part. Beams {\bf 22}, 423 (2004); 
T. Ohkubo, A. G. Zhidkov, T. Hosokai {\it et al.}, Phys. Plasmas {\bf 13}, 033110 (2006); 
 M. Kando, Y. Fukuda, H. Kotaki {\it et al.}, J. Exp. Theor. Phys. {\bf 105}, 916 (2007);
 A. V. Brantov, T. Zh. Esirkepov, M. Kando {\it et al.}, Phys. Plasmas {\bf 15}, 073111 (2008);  
C. G. R. Geddes, K. Nakamura, G. R. Plateau {\it et al.}, Phys. Rev. Lett. {\bf 100}, 215004 (2008); 
J. Faure, C. Rechatin, O. Lundh {\it et al.}, Phys. Plasmas {\bf 17}, 083107 (2010); 
K. Schmid, A. Buck, C. M. S. Sears {\it et al.}, Phys. Rev. ST Accel. Beams {\bf 13}, 091301 (2010);
Y.-C. Ho, T.-S. Hung, C.-P. Yen {\it et al.}, Phys. Plasmas {\bf 18}, 063102 (2011);
A. J. Gonsalves, K. Nakamura, C. Lin {\it et al.}, Nat. Phys. {\bf 7}, 862 (2011); 
Y. Y. Ma, S. Kawata, T. P. Yu {\it et al.}, Phys. Rev. {\bf E 85}, 046403 (2012).

\bibitem{AP}A. I. Akhiezer and R. V. Polovin, J. Exp. Theor. Phys. \textbf{3}, 696 (1956); 
A. I. Akhiezer, I. A. Akhiezer, R. V. Polovin, A. G. Sitenko, and K. N. Stepanov, {\it Plasma
Electrodynamics} (Oxford: Pergamon, 1975).

\bibitem{TWB} S. V. Bulanov, F. Pegoraro, A. M. Pukhov, and A. S. Sakharov, Phys. Rev. Lett. {\bf 78}, 4205 (1997).

\bibitem{PMtV} A. Pukhov and J. Meyer-ter-Vehn, Appl. Phys. B {\bf 74}, 355 (2002).
 
\bibitem{BW1} S. V. Bulanov, Plasma Phys. Controlled Fusion {\bf 48}, 29 (2006)
 
\bibitem{BW2} T. Zh. Esirkepov, T. Kato, and S. V. Bulanov, Phys. Rev. Lett. \textbf{101}, 265001 (2008).

\bibitem{DGBSS} J. M. Dawson, Phys. Rev. \textbf{113}, 383 (1959);
J. F. Drake, Y. C. Lee, K. Nishikawa, and N. L. Tsintsadze, Phys. Rev. Lett. {\bf 36}, 196 (1976);
S. V. Bulanov, L. M. Kovrizhnykh, and A. S. Sakharov, Phys. Rep. {\bf 186}, 1 (1990);
G. Lehmann, E. W. Laedke, and K. H. Spatschek, Phys. Plasmas \textbf{14}, 103109 (2007);
L. M. Gorbunov, A. A. Frolov, E. V. Chizhonkov, and N. E. Andreev, Plasma Phys. Rep. {\bf 36}, 345 (2010);  P. S. Verma, S. Sengupta, and P. Kaw, Phys. Rev. Lett. \textbf{108}, 125005 (2012);
S. S. Bulanov, A. Maksimchuk, C. B. Schroeder, A. G. Zhidkov, E. Esarey, and W. P. Leemans, Phys. Plasmas \textbf{19}, 020702 (2012).


\bibitem{LMF} J. A. Stamper, K. Papadooulos, R. N. Sudan, S. O. Dean, E. A. McLean, and J. M. Dawson, Phys. Rev. Lett. {\bf 26}, 1012 (1971);
U. Wagner, M. Tatarakis, A. Gopal, F. N. Beg, E. L. Clark, A. E. Dangor, R.G. Evans, M. G. Haines, S. P. D. Mangles,
P. A. Norreys, M.-S. Wei, M. Zepf, and K. Krushelnick, Phys. Rev. E {\bf 70}, 026401 (2004);
H. Yoneda, T. Namiki, A. Nishida, R. Kodama, Y. Sakawa, Y. Kuramitsu, T. Morita, K. Nishio, and T. Ide, Phys. Rev. Lett. {\bf 109}, 125004 (2012).

\bibitem{IFE} L. P. Pitaevskii, 
Sov. Phys. JETP {\bf 12}, 1008 (1961); 
J. Deschamps, M. Fitaire, and M. Lagoutte, Phys. Rev. Lett. {\bf 25}, 1330
(1970); Y. Horovitz, S. Eliezer, A. Ludmirsky, Z. Henis, E. Moshe, R. Shpitalnik,
and B. Arad, Phys. Rev. Lett. {\bf 78}, 1707 (1997); 
Y. Horovitz, S. Eliezer, Z. Henis, Y. Paiss, E. Moshe, A. Ludmirsky, M.
Werdiger, B. Arad, and A. Zigler, Phys. Lett. A {\bf 246}, 329 (1998); 
Z. Najmudin, M. Tatarakis, A. Pukhov, E. L. Clark, R. J. Clarke, A. E.
Dangor, J. Faure, V. Malka, D. Neely, M. I. K. Santala, and K.
Krushelnick, Phys. Rev. Lett. {\bf 87}, 215004 (2001); 
V. Yu. Bychenkov and V. T. Tikhonchuk, Laser Part. Beams {\bf 14}, 55
(1996); Z. M. Sheng and J. Meyer-ter-Vehn, Phys. Rev. E {\bf 54}, 1833 (1996); 
N. Naseri, V. Yu. Bychenkov, and W. Rozmus, Phys. Plasmas \textbf{17}, 083109 (2010).

\bibitem{GAA} G. A. Askar'yan, S. V. Bulanov, F. Pegoraro, and A. M. Pukhov, JETP Lett. {\bf 60}, 251 (1994); 
A. Pukhov and J. Meyer-ter-Vehn, Phys. Rev. Lett. {\bf 76}, 3975 (1996).

\bibitem{STRN} T. Katsouleas and J. M. Dawson, Phys. Rev. Lett. {\bf 51}, 392 (1983);

\bibitem{OTHMF} S. V. Bulanov and A. S. Sakharov, JETP Lett. {\bf 44}, 543 (1986); 
V. S. Berezinskii, S. V. Bulanov, V. L. Ginzburg, V. A. Dogiel, V. S. Ptuskin, {\it Astrophysics of cosmic rays} (North Holland, Amsterdam, 1990); 
M. E. Dieckmann, P. K. Shukla, and L. O. C. Drury, ApJ {\bf 675}, 586 (2008);
A. I. Neishtadt, A. V. Artemyev, L. M. Zelenyi, D. L. Vainshtein, JETP Lett. {\bf 89}, 441 (2009).

\bibitem{BINJ} J. Vieira, S. F. Martins, V. B. Pathak, R. A. Fonseca, W. B. Mori, and L. O. Silva, Phys. Rev. Lett. {\bf 106}, 225001 (2011);
J. Vieira, S. F. Martins, V. B. Pathak, R. A. Fonseca, W. B. Mori, and L. O. Silva, Plasma Phys. Control. Fusion {\bf 54}, 124044 (2012). 

\bibitem{CAPILL} Y. Ehrlich, C. Cohen, A. Zigler, J. Krall, P. Sprangle, and E. Esarey, Phys.
Rev. Lett. {\bf 77}, 4186 (1996); N. A. Bobrova, A. A. Esaulov, J.-I. Sakai, P. V. Sasorov, D. J. Spence, A. Butler, 
S. M. Hooker, S. V. Bulanov,  Phys. Rev. E {\bf 65}, 016407 (2002); T. Kameshima, H. Kotaki, M. Kando, I. Daito, K. Kawase, 
Y. Fukuda, L. M. Chen, T. Homma, S. Kondo, T. Zh. Esirkepov, N. A. Bobrova, P. V. Sasorov, and S. V. Bulanov,
Phys. Plasmas {\bf 16}, 093101 (2009).

\bibitem{THBf} T. Hosokai, K. Kinoshita, A. Zhidkov, A. Maekawa, A. Yamazaki, and M. Uesaka, Phys. Rev. Lett. {\bf 97}, 075004 (2006); 
T. Hosokai, A. Zhidkov, A. Yamazaki, Y. Mizuta, M. Uesaka, and R. Kodama, Appl. Phys. Lett. {\bf 96}, 121501 (2010).

\bibitem{UHWW} P. Shukla, Phys. Scr. {\bf T52}, 73 (1994);
P. Shukla, Phys. Plasmas {\bf 6}, 1363 (1999);
A. Holkundkar, G. Brodin, and M. Marklund, Phys. Rev. E {\bf 84}, 036409 (2011).

\bibitem{UHECR} P. Chen, T. Tajima, and Y. Takahashi, Phys. Rev. Lett. {\bf 89}, 161101 (2002);
F.-Y. Chang, P. Chen, G.-L. Lin, R. Noble, and R. Sydora, Phys. Rev. Lett. {\bf 102}, 111101 (2009).

\bibitem{RD} R. C. Davidson, {\it Methods in Nonlinear Plasma Theory} (Academic, New York, 1972).

\bibitem{TBI} S. V. Bulanov, T. Zh. Esirkepov, M. Kando, J. Koga, A. S. Pirozhkov, T. Nakamura, S. S. Bulanov, C. B. Schroeder, E. Esarey, F. Califano, and F. Pegoraro, Phys. Plasmas {\bf 19}, 113102 (2012).

\bibitem{Schmit2012} P. F. Schmit and N. J. Fisch, Phys. Rev. Lett. {\bf 109}, 255003 (2012).

\bibitem{Bourdier2012} A. Bourdier, G. Girard, S. Rassou, X. Davoine, and M. Drouin, J. Mod. Phys. {\bf 3}, 1983 (2012).

\bibitem{BetaX} A. Rousse {\it et al.}, Phys. Rev. Lett. {\bf 93}, 135005 (2004); 
K. Ta Phuoc, F. Burgy, J.-P. Rousseau, V. Malka, A. Rousse, R. Shah, D. Umstadter, A. Pukhov, and S. Kiselev, Phys. Plasmas {\bf 12}, 023101 (2005). 

\bibitem{BetaTheory1} E. Esarey, B. A. Shadwick, P. Catravas, and W. P. Leemans, Phys. Rev. E {\bf 65},056505 (2002);  I. Kostyukov, S. Kiselev, and A. Pukhov, Phys. Plasmas {\bf 10}, 4818 (2003).

\bibitem{BetaTheory2} S. V. Bulanov, M. Yamagiwa, T. Zh. Esirkepov, J. K. Koga,
M. Kando, Y. Ueshima, and K. Saito, Phys. Plasmas {\bf 12}, 073103 (2005).

\bibitem{BetaImage} K. T. Phuoc, S. Corde, R. Shah, F. Albert, R. Fitour, J. P. Rousseau,
F. Burgy, B. Mercier, and A. Rousse, Phys. Rev. Lett. {\bf 97}, 225002 (2006);
S. Corde, K. Ta Phuoc, G. Lambert, R. Fitour, V. Malka, A. Rousse, A. Beck and E. Lefebvre, Rev. Mod. Phys. {\bf 85}, 1 (2013) .

\bibitem{REMP} T. Zh. Esirkepov, Comput. Phys. Commun. {\bf 135}, 144 (2001).

\end{thebibliography}
\end{document}